\documentclass{emulateapj}
\usepackage{amssymb}
\usepackage{amsmath}

\newcommand{\mbh}{\ensuremath{M_{\rm{BH}}}\,}
\newcommand{\er}{\ensuremath{\lambda\,}}

\received{August 2, 2017}
\revised{August 31, 2017}
\accepted{September 19, 2017}

\shorttitle{Near-IR spectroscopy of luminous LoBAL quasars at $1<z<2.5$}
\shortauthors{Schulze et al.}

\begin{document}

\title{Near-IR spectroscopy of luminous LoBAL quasars at $1<z<2.5$}

\author{Andreas Schulze\altaffilmark{1,11}, Malte Schramm\altaffilmark{1}, Wenwen Zuo\altaffilmark{2}, Xue-Bing Wu\altaffilmark{3,4}, Tanya Urrutia\altaffilmark{5}, Jari Kotilainen\altaffilmark{6,7}, Thomas Reynolds\altaffilmark{7},  Koki Terao\altaffilmark{8}, Tohru Nagao\altaffilmark{9}, Hideyuki Izumiura\altaffilmark{10}}
\email{E-mail: andreas.schulze@nao.ac.jp}

\altaffiltext{1}{National Astronomical Observatory of Japan, Mitaka, Tokyo 181-8588, Japan}
\altaffiltext{2}{Shanghai Astronomical Observatory, Shanghai 200030, China}
\altaffiltext{3}{Department of Astronomy, Peking University, Yi He Yuan Lu 5, Hai Dian District, Beijing 100871, China}
\altaffiltext{4}{Kavli Institute for Astronomy and Astrophysics, Peking University, Beijing 100871, China}
\altaffiltext{5}{Leibniz-Institut f\"ur Astrophysik Potsdam (AIP), An der Sternwarte 16, 14482 Potsdam, Germany}
\altaffiltext{6}{Finnish Centre for Astronomy with ESO (FINCA), University of Turku, V\"ais\"al\"antie 20, FI-21500 Piikki\"o, Finland}
\altaffiltext{7}{Tuorla Observatory, Department of Physics and Astronomy, University of Turku, V\"ais\"al\"antie 20, FI-21500 Piikki\"o, Finland}
\altaffiltext{8}{Department of Physics, Ehime University, Bunkyo-cho, 2-5, Matsuyama, 790-8577, Japan}
\altaffiltext{9}{Research Center for Space and Cosmic Evolution, Ehime University, Bunkyo-cho 2-5, Matsuyama, 790-8577, Japan}
\altaffiltext{10}{Okayama Astrophysical Observatory, National Astronomical Observatory of Japan, National Institutes of Natural Sciences,
3037-5 Honjo, Kamogata, Asakuchi, Okayama 719-0232, Japan}
\altaffiltext{11}{EACOA Fellow}



\begin{abstract}
We present near-IR spectroscopy of 22 luminous low-ionization broad absorption line quasars  (LoBAL QSOs) at  redshift $1.3<z<2.5$, with 12 objects at $z\sim1.5$ and 10 at $z\sim2.3$. The spectra cover the rest-frame H$\alpha$ and H$\beta$ line regions, allowing us to obtain robust black hole mass estimates based on the broad H$\alpha$ line. We use these data, augmented by a lower-redshift sample from the Sloan Digital Sky Survey, to test the proposed youth scenario for LoBALs, which suggests that LoBALs constitute an early short-lived evolutionary stage of quasar activity, by probing for any difference in their masses, Eddington ratios, or rest-frame optical spectroscopic properties compared to normal quasars. In addition, we construct the UV to mid-IR spectral energy distributions (SEDs) for the LoBAL sample and a matched non-BAL quasar sample. We do not find any statistically significant difference between LoBAL QSOs and non-BAL QSOs in their black hole mass or Eddington ratio distributions. The mean UV to mid-IR SED of the LoBAL QSOs is consistent with non-BAL QSOs, apart from their stronger reddening. At $z>1$ there is no clear difference in their optical emission line properties. We do not see particularly weak [\ion{O}{3}] or strong \ion{Fe}{2} emission. The LoBAL QSOs do not show a stronger prevalence of ionized gas outflows as traced by the [\ion{O}{3}] line, compared to normal QSOs of similar luminosity. We conclude that the optical$-$MIR properties of LoBAL QSOs are consistent with the general quasar population and do not support them to constitute a special phase of active galactic nucleus evolution.
\end{abstract}


\section{Introduction} \label{sec:intro}
Broad Absorption Line quasars (BAL QSOs) are an important, yet still not well understood sub-class of the quasar population, which shows evidence for energetic outflows via the presence of strong blueshifted broad absorption lines with velocities up to $0.2c$ \citep[e.g.][]{Foltz:1983,Weymann:1991,Jannuzi:1996}. AGN outflows are of fundamental importance for our understanding of the AGN feedback mechanism, which is thought to be able to quench star formation and self-regulate the growth of the supermassive black hole (SMBH) and of its host galaxy \citep[e.g.][]{Silk:1998,DiMatteo:2005,Fabian:2012}. They therefore might play a role in establishing the observed relations between SMBH mass and the properties of its host galaxy \citep[e.g.][]{Kormendy:2013}. 

BAL systems are present in $\sim15$\% of quasars in optically selected samples \citep{Hewett:2003,Gibson:2009}, but might have an intrinsic fraction as high as $\sim40$\% \citep{Allen:2011}.  They are identified by the presence of blueshifted absorption mainly in high-ionization lines, such as \ion{C}{4} and \ion{Si}{4}. A small fraction of BALs, about $\sim15$\%,  in addition also shows broad absorption in  low-ionization ions like \ion{Mg}{2}  and \ion{Al}{3}. These are called  LoBAL QSOs, while the former are termed HiBAL QSOs. Even less common are FeLoBALs, which in addition to the low-ionization BAL also show absorption troughs in the metastable \ion{Fe}{2} line \citep{Hazard:1987,Becker:1997,Hall:2002}. There are also a few rare cases known which show broad absorption features even in the Balmer lines \citep[e.g.][]{Aoki:2006,Hall:2007}. 

While BAL systems represent the most extreme forms of intrinsic quasar absorption systems, quasar outflows are also detected in narrow absorption lines (NAL) and mini-BALs, with line widths  $<2000$~km s$^{-1}$ to a few hundred km s$^{-1}$ \citep[e.g.][]{Hamann:1997,Vestergaard:2003,Misawa:2007}. The mass outflows seen in BAL QSOs are thought to be launched as radiation-driven disc winds \citep[e.g.][]{Proga:2000,Proga:2004}. 

There are two main interpretations proposed to explain the BAL phenomenon. The first is an orientation scenario, which argues that most quasars have a BAL wind, but their broad absorption line region (BALR) covering fraction is low, so the quasar can be observed as a BAL QSO only along a particular line-of-sight. The second interpretation is an evolution scenario, where a BAL QSO represents a particular stage in quasar evolution, possibly with a high BALR covering fraction. So the quasar is observed as a BAL along most lines-of-sight if in this particular evolutionary stage.

For HiBAL QSOs the orientation scenario is the most plausible interpretation. This is observationally supported by their similar continuum and emission line properties \citep{Weymann:1991, Reichard:2003}, spectropolarimetric observations \citep{Ogle:1999,Schmidt:1999} and their similar spectral energy distribution \citep{Willott:2003,Gallagher:2007,CaoOrjales:2012}. Furthermore, HiBALs often show time variability in their absorption strength, including the disappearance and re-emergence of the \ion{C}{4} BAL \citep[e.g.][]{FilizAk:2012,McGraw:2017}.

On the other hand, for LoBAL QSOs and especially for FeLoBAL QSOs an evolution scenario has been suggested, in which they constitute an early stage of quasar evolution \citep{Boroson:1992b,Voit:1993,Becker:2000}. In this picture, LoBAL QSOs are young AGN in a short-lived transition phase between an ultra-luminous infrared galaxy (ULIRG) and a normal unobscured quasar. It is thought that a merger induced young QSO, enclosed before by a dust rich cocoon and observed as a ULIRG, is ignited and blows off their dust envelope by a powerful wind, accreting at a high rate. This quasar wind may provide AGN feedback to quench star-formation in their host galaxy \citep{Farrah:2012,Faucher:2012}.

There are a few pieces of observational evidence to support this picture.
LoBAL QSOs show strong reddening, due to high levels of dust extinction $E(B-V)\sim0.14$ \citep{Sprayberry:1992,Brotherton:2001,Reichard:2003, Gibson:2009}. Many of them will be therefore missed in surveys using optical color selection. Indeed, while their fraction in optical quasar samples is low ($\sim1$\%), LoBALs are much more common in near-IR selected samples \citep{Urrutia:2009,Dai:2012}.
(Fe)LoBALs are often found to be associated with high FIR luminosities and high star formation rates \citep{Canalizo:2002,Farrah:2007,Farrah:2010}. However, more recent studies did not find evidence for a significantly different level of star formation in LoBAL QSOs compared to normal quasars \citep{Lazarova:2012,Violino:2016}. Several LoBALs/FeLoBALs show signatures of interactions or major mergers \citep{Canalizo:2002,Gregg:2002}, but currently studies suffer from small number statistics. Furthermore low-$z$ LoBAL QSOs show differences in their rest-frame optical spectra, which point to LoBAL QSOs as a special quasar sub-class rather than an orientation effect, for example having on average weak [\ion{O}{3}] and strong \ion{Fe}{2} emission \citep{Boroson:1992b,Zhang:2010,Runnoe:2013}. 
They also show variability in their absorption strength \citep{Hall:2011,FilizAk:2014,Rafiee:2016}, though based on current samples not to the extent of full disappearance of the low-ionization troughs.

One implication of the young QSO scenario for LoBALs implies that they should have on average high Eddington ratios \citep{Zubovas:2013}. Testing this prediction requires the measurement of SMBH masses for a representative LoBAL QSO sample. 
Previous studies on LoBAL QSOs or red quasars at $z<1$ indeed found tentative evidence for high accretion rates \citep{Boroson:2002, Zhang:2010, Urrutia:2012}, while \citet{Runnoe:2013} found for a small sample of radio-selected BAL QSOs (mainly LoBALs) that they are not predominantly accreting at or above the Eddington limit.
At $z>1$ this has not been probed yet.
 
For broad line AGN, black hole masses can be estimated from single-epoch spectroscopy via the established 'virial method' \citep[e.g.][]{Vestergaard:2006}, using their broad emission lines, like H$\beta$, H$\alpha$, \ion{Mg}{2} and \ion{C}{4}, and the continuum luminosity. However, for LoBAL QSOs at $z>1$ broad \ion{Mg}{2} and/or \ion{C}{4} are the only lines available in optical spectroscopy, but they are not suited for SMBH mass \mbh estimation, due to the strong reddening and the significant absorption either directly affecting the line or the neighboring continuum. Therefore BALQSOs are usually excluded from SMBH mass studies. 
The broad H$\alpha$ or H$\beta$ lines are much less affected by reddening and absorption, and thus provide the most robust estimator of \mbh in LoBAL QSOs.
However  beyond $z\sim1$ their observation requires near-IR spectroscopy, making reliable SMBH mass estimates of LoBAL QSOs beyond this redshift currently rare, despite the fact that AGN activity and the major merger rate are much higher at these redshifts, which makes the redshift range $1<z<3$ a crucial period in the  black hole growth history.

We here present near-IR spectroscopy of  H$\alpha$ and H$\beta$ for a well defined sample of 22 LoBAL QSOs at $1.3<z<2.5$. We use these to estimate their SMBH masses and Eddington ratios as well as to investigate their rest-frame optical spectral properties and thereby test the proposed evolutionary scenario for this QSO population.

Throughout this paper we assume a Hubble constant of $H_0 = 70$ km s$^{-1}$ Mpc$^{-1}$ and cosmological density parameters $\Omega_\mathrm{m} = 0.3$ and $\Omega_\Lambda = 0.7$. Near-IR 2MASS magnitudes are given in the Vega system.

\begin{deluxetable*}{lcccccccccccccc}
\tabletypesize{\scriptsize}
\tablecaption{Sample summary}
\tablewidth{18cm}
\tablehead{
\colhead{Name} & \colhead{R.A. (J2000)} & \colhead{Decl. (J2000)} & \colhead{$z_{\rm{NIR}}$} & \colhead{$z_{\rm{HW}}$}& \colhead{BI(\ion{Mg}{2})} & \colhead{BI(\ion{Al}{3})} &  \colhead{$H$} & \colhead{$K$} &  \colhead{Inst.(H$\beta$)} & \colhead{Inst.(H$\alpha$)} &  \colhead{Type} 
}
\startdata
SDSS J0033+0632 & 00:33:35.638 & +06:32:07.58  & 1.502 &  1.505 &  188.7 &  457.3 & 15.785 & 15.748 & TSPEC & TSPEC &  Lo\\ 
SDSS J0132$-$0046 & 01:32:45.302 & -00:46:10.01  & 1.469 &  1.475 &  147.6 &  342.4 & 16.539 &         & TSPEC & TSPEC &  Lo\\ 
SDSS J0859+4239 & 08:59:10.400 & +42:39:11.38 &  1.497 &  1.499 & 5212.0 &  496.2 & 15.381 & 15.853 & TSPEC & TSPEC &  Lo\\ 
SDSS J0952+0257 & 09:52:32.212 & +02:57:28.39 &  1.358 &  1.358 & 1034.5 &  648.0 & 15.493 & 15.424 & TSPEC & TSPEC &  Lo\\ 
SDSS J0957+4406 & 09:57:21.361 & +44:06:42.91 &  1.459 &  1.468 &   90.0 &  381.5 & 15.665 & 15.169 & TSPEC & TSPEC &  Lo\\ 
SDSS J1019+0225 & 10:19:27.371 & +02:25:21.44 &  1.364 &  1.364 & 4648.4 &    0.0 & 15.215 & 15.114 & TSPEC & TSPEC &  Lo\\ 
SDSS J1128+0623 & 11:28:51.837 & +06:23:15.38 &  1.513 &  1.497 &   67.3 &    0.0 & 14.754 & 14.438 & TSPEC & TSPEC &  Lo\\ 
SDSS J1440+3710 & 14:40:02.245 & +37:10:58.52 &  1.401 &  1.414 & 1540.7 &    0.0 & 15.316 & 14.761 &   & ISLE &  FeLo\\ 
SDSS J1448+0424 & 14:48:42.451 & +04:24:03.12 &  1.539 &  1.546 &   64.3 &  126.1 & 14.482 & 14.402 & TSPEC & TSPEC &  Lo\\ 
SDSS J1508+6055 & 15:08:48.805 & +60:55:51.93 &  1.529 &  1.532 & 3068.7 &    0.0 & 15.241 & 14.756 & NOTCam & ISLE &  FeLo\\ 
SDSS J1511+4905 & 15:11:13.846 & +49:05:57.37 &  1.368 &  1.361 &  369.8 & 1264.5 & 14.606 & 14.284 &   & IRCS &  Lo\\ 
SDSS J1556+3517 & 15:56:33.783 & +35:17:57.39 &  1.501 &  1.495 & 9926.2 &    0.0 & 14.905 & 14.787 & TSPEC & TSPEC &  FeLo\\ 
\noalign{\smallskip} \hline \noalign{\smallskip}
SDSS J0841+2005 & 08:41:33.153 & +20:05:25.81 &  2.345 &  2.276 & $-$ & 11540.6 & 14.411 & 13.620 & NOTCam & ISLE &  FeLo\\ 
SDSS J0943$-$0100 & 09:43:38.218 & -01:00:19.33 &  2.368 &  2.376 & $-$ &  395.0 &   & 14.947 & NOTCam & NOTCam &  Lo\\ 
SDSS J1011+5155 & 10:11:08.895 & +51:55:53.82 &  2.472 &  2.465 & $-$ & 3182.4 & 15.979 & 15.245 &   & NOTCam &  Lo\\ 
SDSS J1019+4108 & 10:19:12.850 & +41:08:07.41 &  2.471 &  2.460 & $-$ & 6287.6 & 15.828 & 14.685 &   & NOTCam &  Lo\\ 
SDSS J1028+5110 & 10:28:50.317 & +51:10:53.11 &  2.418 &  2.426 & $-$ &  128.7 & 15.591 & 14.660 & NOTCam & NOTCam &  FeLo\\ 
SDSS J1132+0104 & 11:32:12.920 & +01:04:41.35 &  2.377 &  2.328 & $-$ &  884.0 & 16.254 & 15.177 &   & NOTCam &  Lo\\ 
SDSS J1134+3238 & 11:34:24.642 & +32:38:02.45 &  2.454 &  2.461 & $-$ & 3178.3 & 14.844 & 13.987 & NOTCam & ISLE &  FeLo\\ 
SDSS J1516+0029 & 15:16:36.786 & +00:29:40.51 &  2.252 &  2.251 & $-$ &  597.7 & 15.701 & 15.007 &   & NOTCam &  Lo\\ 
SDSS J1554+2218 & 15:54:33.131 & +22:18:42.08 &  2.410 &  2.418 & $-$ &   41.3 & 15.810 & 14.731 & NOTCam & NOTCam &  Lo\\ 
SDSS J1709+6303 & 17:09:30.996 & +63:03:57.13 &  2.380 &  2.407 & $-$ &  482.7 & 15.551 & 14.624 & NOTCam & ISLE &  Lo\\
\enddata
\tablecomments{$z_{\rm{NIR}}$ is the redshift measured from the peak of either H$\alpha$ or [\ion{O}{3}]; $z_{\rm{HW}}$ is the improved SDSS redshift from \citet{Hewett:2010}; BI(\ion{Mg}{2}) and BI(\ion{Al}{3}) are the balnicity indices for these broad lines as measured by \citet{Allen:2011};  $H_{\rm{2MASS}}$ and $K_{\rm{2MASS}}$ are the 2MASS magnitudes taken from \citet{Schneider:2010}; Inst. gives the instrument used for spectroscopy of either $H\alpha$ or H$\beta$ (note that only TSPEC spectroscopy covers $H\alpha$ and H$\beta$ simultaneously); Type indicates if the object is a regular LoBAL (Lo) or an obvious FeLoBAL (FeLo).}
\label{tab:sample}
\end{deluxetable*}

\section{Sample and Observations}
\subsection{Sample selection}
Our sample is drawn from the BAL QSO catalog from \citet{Allen:2011}. They measured BAL properties for the high-ionization lines \ion{Si}{4}$\lambda$1400 and \ion{C}{4}$\lambda$1550 and the low-ionization lines \ion{Al}{3}$\lambda$1860 and \ion{Mg}{2}$\lambda$2800 from quasar spectra in the SDSS DR6 spectroscopic survey \citep{Schneider:2007, Schneider:2010}. A BAL QSO in their sample is defined as having a non-zero balnicity index (BI), where the BI is defined following \citet{Weymann:1991} and measures the presence of a continuous broad absorption feature below a threshold of 0.9 in respect to the normalized continuum.

Using this definition, their sample contains 368 LoBAL QSOs, identified by an $\mathrm{BI}>0$ either in  \ion{Mg}{2} or \ion{Al}{3}. We are selecting our targets in two redshift windows from this LoBAL QSO sample. These are chosen such that H$\alpha$ falls well within the atmospheric window not strongly affected by telluric absorption bands in either $H$-band or $K$-band. In addition, for most targets H$\beta$ falls into $J$-band or $H$-band respectively. Specifically, we target:

(1) the redshift range $1.32<z<1.60$, BI(\ion{Mg}{2})$>0$ and a 2MASS $H$-band magnitude $H<16.7$~mag, giving a sample of 23 targets. We removed SDSSJ014349.15+002128.3 from the sample, since this object is not a true LoBAL QSO but a superposition of a normal QSO with intervening absorption by a foreground star (Wenjuan Liu, private communication). These 22 LoBAL QSOs form our initial $z\sim1.5$ sample. We obtained near-IR spectroscopy for 12 of these LoBAL targets. For 10 of these we have spectroscopy for both H$\alpha$ and H$\beta$.

(2) the redshift range $2.2<z<2.5$, BI(\ion{Al}{3})$>0$ and a 2MASS $K$-band magnitude $K<15.3$~mag. The relatively bright luminosity cut is motivated by the smaller size telescopes used for near-IR spectroscopy of the sample in this redshift bin to ensure acceptable signal-to-noise in the spectra. Focusing on these bright LoBALs gives a sample of 11 targets, forming our initial $z\sim2.3$ sample. We obtained $K$-band spectra for 10 of them, covering  H$\alpha$, and additionally $H$-band spectra covering H$\beta$ for 7 of these, where 6 have acceptable S/N and are used here. For 9 of the $z\sim2.3$ LoBALs BOSS spectra are available which cover also the \ion{Mg}{2} line at these redshifts, contrary to SDSS-I/II spectra. The presence of a clear absorption trough also in \ion{Mg}{2} can be confirmed from the BOSS spectra.

The basic sample properties are provided in Table~\ref{tab:sample}. We indicate clear cases of FeLoBALs in our sample. In total, we identified three FeLoBALs in the $z\sim1.5$ sample and another three in the $z\sim2.3$ sample, based on visual inspection. SDSS~J0841+2005 is known as an FeLoBAL which shows very strong changes in its absorption systems \citep{Rafiee:2016,Stern:2017}.

In the $z\sim1.5$ LoBAL sample we have discovered two cases with strong intrinsic Balmer absorption lines in H$\alpha$ and H$\beta$ through our observations. Intrinsic Balmer absorption in quasar spectra is a rare phenomenon, with only a handful of cases known so far \citep[e.g.][]{Aoki:2006,Hall:2007,Zhang:2015}. We include these objects here in our statistical LoBAL study and refer a more detailed discussion to a separate paper (Schulze et al. in prep.).

\subsection{Observations and Data reduction}
For our $z\sim1.5$ sample we used the near-IR spectrograph TripleSpec \citep{Wilson:2004} at the Palomar Hale 200 inch telescope to observe 9 of our $z\sim1.5$ targets in January 2014 under good conditions. TripleSpec provides simultaneous coverage from 1.0 $\mu$m to 2.4 $\mu$m at a spectral resolution of R$\sim2700$. A slit width of 1\arcsec\ was used. Total exposure times varied between 40 and 60~min.
Observations for 3 other quasars were obtained using ISLE on the 1.88m telescope at Okayama Astrophysical Observatory (OAO), NOTCam on the 2.56m Nordic Optical Telescope (NOT) and IRCS on the 8.2m Subaru telescope \citep{Kobayashi:2000}.

For the $z\sim2.3$ sample we mainly used the 2.56m NOT, supplemented by the OAO 1.88m telescope for three targets. Using NOTCam on the NOT, we obtained low-resolution R$=2500$ spectroscopy in either $J$, $H$ or $K$ band with a slit width of 0.6\arcsec.
Observations were carried out during two runs in March 2016 and March 2017 under very good conditions with an average seeing of 0.7\arcsec. Typical exposure times range between 30-60~min.

Observations at OAO were obtained during several runs from 2015-2017 under mostly poor conditions with varying seeing between 1-2.5\arcsec. Due to the seeing limitations we increased the slit width to 2\arcsec\ to avoid significant slit losses, leading to a reduced resolution of R$\sim$1000. Exposure times per target are between 1-2~hours. For each quasar a full calibration set (including dome flats and Xe and Ar arc lamps) was observed together with a telluric standard star at similar airmass and position either before or after the quasar.
We performed an ABBA dither pattern along the slit to improve the sky subtraction for all targets at each of these telescopes.

The data reduction for the spectroscopic data from TripleSpec is carried out using the modified IDL-based Spextool3 package
\citep{Cushing:2004}, as described in \citet{Zuo:2015}. This involved flat field correction, sky background subtraction, wavelength calibration and telluric correction. The telluric correction is based on several A0V stars observed each night. The data reduction of the spectra from the other facilities was performed using the IRAF software following the standard reduction steps for sky-subtraction, flat-fielding and telluric correction. We extracted the 1D spectrum and performed a wavelength calibration using either Ar or Xe arc-lamp.

We did not perform spectro-photometric flux calibration but tied the absolute flux calibration to their 2MASS magnitudes. Simultaneous NIR $K$-band observations of three of our targets using the Wide-Field Imager mounted on the 91 cm telescope at OAO showed that the current NIR photometry is fully consistent with their 2MASS photometry. All spectra are corrected for galactic extinction \citep{Cardelli:1989, Schlegel:1998}.

\begin{figure*}
\centering
\includegraphics[width=8.6cm,clip]{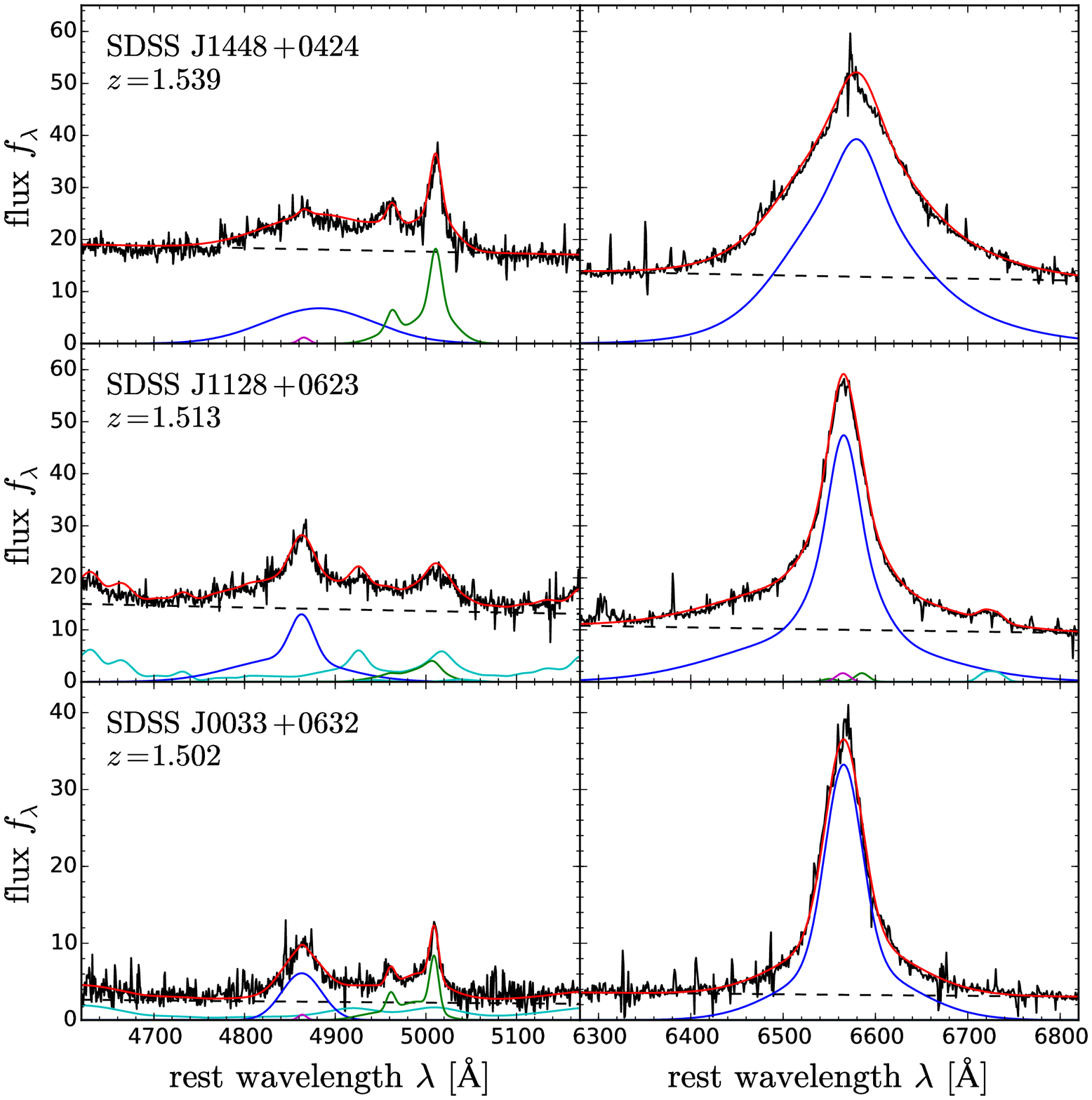} \hspace{0.5cm}
\includegraphics[width=8.6cm,clip]{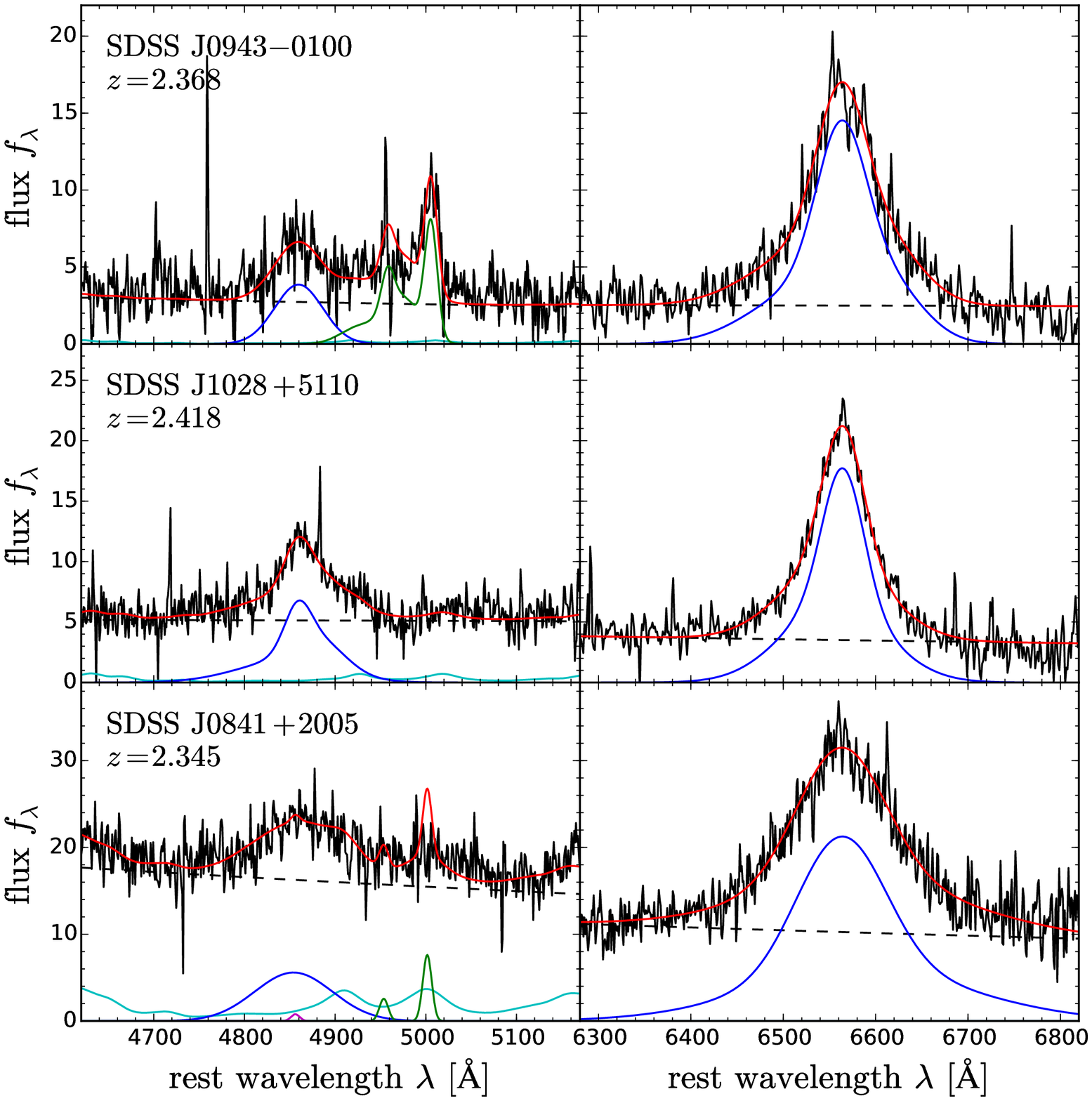}
\caption{We show example spectra from the $z\sim1.5$ sample (left panel) and the $z\sim2.3$ sample (right panel) with their best fit continuum+emission line model, including a power-law continuum (black dashed line), an \ion{Fe}{2} template (cyan), a multi-Gauss model for the broad Balmer lines (blue) and [\ion{O}{3}] (green) and a narrow Balmer line (magenta), [\ion{N}{2}] (green) and [\ion{S}{2}] (cyan) component if justified.}
\label{fig:specex} 
\end{figure*}

\begin{deluxetable*}{lcccccccccc}
\tabletypesize{\scriptsize}
\tablecaption{Spectral measurements and derived black hole properties}
\tablewidth{18cm}
\tablehead{
\colhead{Name} & \colhead{FWHM$_{\rm{H}\alpha}$} & \colhead{$\log L_{\rm{H}\alpha}$} & \colhead{FWHM$_{\rm{H}\beta}$} & \colhead{$\log L_{\rm{H}\beta}$} & \colhead{$\log L_{5100}$} & \colhead{$\log L_{\rm{bol}}$} &  \colhead{$\log \mbh$} & \colhead{$\log \er$} \\
\colhead{} & \colhead{(km\ s$^{-1}$)} & \colhead{(erg s$^{-1}$)} & \colhead{(km\ s$^{-1}$)} & \colhead{(erg s$^{-1}$)} & \colhead{(erg s$^{-1}$)} & \colhead{(erg s$^{-1}$)} &  \colhead{($M_\odot$)} & \colhead{} 
}
\startdata
SDSS J0033+0632 & 2475 $\pm$ 29 &  44.95 $\pm$   0.01 & 3165 $\pm$ 107 &  44.08 $\pm$   0.02 &  45.60 $\pm$   0.03 &  47.06 $\pm$   0.01 &   8.96 $\pm$   0.01 &  $-$0.01 $\pm$   0.01 \\ 
SDSS J0132$-$0046 & 4973 $\pm$ 130 &  44.42 $\pm$   0.01 & 7611 $\pm$ 1745 &  43.37 $\pm$   0.06 &  45.20 $\pm$   0.01 &  46.53 $\pm$   0.01 &   9.35 $\pm$   0.02 &  $-$0.93 $\pm$   0.02 \\ 
SDSS J0859+4239 & 8317 $\pm$ 1145 &  45.02 $\pm$   0.01 & 7895 $\pm$ 500 &  44.11 $\pm$   0.03 &  45.97 $\pm$   0.01 &  47.14 $\pm$   0.01 &  10.11 $\pm$   0.18 &  $-$1.09 $\pm$   0.18 \\ 
SDSS J0952+0257 & 3048 $\pm$ 18 &  44.85 $\pm$   0.01 & 3594 $\pm$ 184 &  43.95 $\pm$   0.02 &  46.03 $\pm$   0.01 &  46.97 $\pm$   0.01 &   9.11 $\pm$   0.01 &  $-$0.25 $\pm$   0.01 \\ 
SDSS J0957+4406 & 7880 $\pm$ 126 &  44.64 $\pm$   0.01 & 5853 $\pm$ 473 &  43.78 $\pm$   0.04 &  46.02 $\pm$   0.01 &  46.75 $\pm$   0.01 &   9.88 $\pm$   0.02 &  $-$1.24 $\pm$   0.02 \\ 
SDSS J1019+0225 & 6440 $\pm$ 3573 &  45.01 $\pm$   0.01 & 8545 $\pm$ 919 &  44.07 $\pm$   0.06 &  45.94 $\pm$   0.01 &  47.12 $\pm$   0.01 &   9.87 $\pm$   0.35 &  $-$0.86 $\pm$   0.35 \\ 
SDSS J1128+0623 & 2574 $\pm$ 61 &  45.22 $\pm$   0.01 & 2580 $\pm$ 208 &  44.54 $\pm$   0.02 &  46.39 $\pm$   0.01 &  47.33 $\pm$   0.01 &   9.13 $\pm$   0.02 &   0.10 $\pm$   0.02 \\ 
SDSS J1440+3710 & 4584 $\pm$ 104 &  44.57 $\pm$   0.01 & $-$ &   $-$ &   $-$ &  46.68 $\pm$   0.01 &   9.35 $\pm$   0.02 &  $-$0.77 $\pm$   0.02 \\ 
SDSS J1448+0424 & 5531 $\pm$ 158 &  45.36 $\pm$   0.01 & 8664 $\pm$ 537 &  44.60 $\pm$   0.03 &  46.53 $\pm$   0.01 &  47.47 $\pm$   0.01 &   9.90 $\pm$   0.03 &  $-$0.54 $\pm$   0.03 \\ 
SDSS J1508+6055 & 3601 $\pm$ 1957 &  44.98 $\pm$   0.01 & 4694 $\pm$ 373 &  44.18 $\pm$   0.03 &  46.22 $\pm$   0.01 &  47.09 $\pm$   0.01 &   9.32 $\pm$   0.33 &  $-$0.34 $\pm$   0.33 \\ 
SDSS J1511+4905 & 2792 $\pm$ 1086 &  45.10 $\pm$   0.14 & $-$ &   $-$ &   $-$ &  47.21 $\pm$   0.14 &   9.14 $\pm$   0.24 &  $-$0.04 $\pm$   0.24 \\ 
SDSS J1556+3517 & 4688 $\pm$ 227 &  45.17 $\pm$   0.01 & 4153 $\pm$ 176 &  44.05 $\pm$   0.03 &  46.16 $\pm$   0.01 &  47.29 $\pm$   0.01 &   9.66 $\pm$   0.04 &  $-$0.48 $\pm$   0.04 \\ 
\noalign{\smallskip} \hline \noalign{\smallskip}
SDSS J0841+2005 & 6107 $\pm$ 181 &  45.70 $\pm$   0.01 & 6225 $\pm$ 1019 &  44.93 $\pm$   0.07 &  47.04 $\pm$   0.01 &  47.82 $\pm$   0.01 &  10.15 $\pm$   0.02 &  $-$0.45 $\pm$   0.02 \\ 
SDSS J0943$-$0100 & 3804 $\pm$ 165 &  45.35 $\pm$   0.01 & 3849 $\pm$ 3783 &  44.58 $\pm$   0.12 &  46.27 $\pm$   0.02 &  47.46 $\pm$   0.01 &   9.55 $\pm$   0.04 &  $-$0.20 $\pm$   0.04 \\ 
SDSS J1011+5155 & 5186 $\pm$ 296 &  45.02 $\pm$   0.03 & $-$ &   $-$ &   $-$ &  47.14 $\pm$   0.03 &   9.68 $\pm$   0.05 &  $-$0.65 $\pm$   0.05 \\ 
SDSS J1019+4108 & 3157 $\pm$ 308 &  45.08 $\pm$   0.05 & $-$ &   $-$ &   $-$ &  47.19 $\pm$   0.05 &   9.25 $\pm$   0.09 &  $-$0.16 $\pm$   0.09 \\ 
SDSS J1028+5110 & 3109 $\pm$ 334 &  45.38 $\pm$   0.02 & 3562 $\pm$ 386 &  44.91 $\pm$   0.03 &  46.61 $\pm$   0.02 &  47.49 $\pm$   0.02 &   9.38 $\pm$   0.09 &   0.01 $\pm$   0.09 \\ 
SDSS J1132+0104 & 4656 $\pm$ 399 &  45.16 $\pm$   0.03 & $-$ &   $-$ &   $-$ &  47.28 $\pm$   0.03 &   9.65 $\pm$   0.08 &  $-$0.48 $\pm$   0.08 \\ 
SDSS J1134+3238 & 6242 $\pm$ 178 &  45.57 $\pm$   0.01 & 4682 $\pm$ 516 &  44.63 $\pm$   0.06 &  46.94 $\pm$   0.01 &  47.68 $\pm$   0.01 &  10.11 $\pm$   0.03 &  $-$0.54 $\pm$   0.03 \\ 
SDSS J1516+0029 & 3826 $\pm$ 1571 &  45.25 $\pm$   0.06 & $-$ &   $-$ &   $-$ &  47.37 $\pm$   0.06 &   9.51 $\pm$   0.28 &  $-$0.25 $\pm$   0.28 \\ 
SDSS J1554+2218 & 3762 $\pm$ 353 &  45.34 $\pm$   0.02 & 6300 $\pm$ 1080 &  44.49 $\pm$   0.05 &  46.55 $\pm$   0.01 &  47.46 $\pm$   0.02 &   9.54 $\pm$   0.07 &  $-$0.19 $\pm$   0.07 \\ 
SDSS J1709+6303 & 4551 $\pm$ 720 &  45.10 $\pm$   0.06 & 8329 $\pm$ 1034 &  44.81 $\pm$   0.06 &  46.60 $\pm$   0.02 &  47.21 $\pm$   0.06 &   9.59 $\pm$   0.14 &  $-$0.49 $\pm$   0.14 \\
\enddata
\tablecomments{$L_{\rm{bol}}$, \mbh and \er have been derived from the broad H$\alpha$ line as discussed in the text.}
\label{tab:prop}
\end{deluxetable*}
 
 \section{Results}  \label{sec:results}
 \subsection{Spectral Measurements}  \label{sec:fitting}
Spectral measurements are obtained from fitting the spectral regions around the broad H$\alpha$ and H$\beta$ line with a continuum+line profile model. Our procedure for continuum fitting and emission line modeling is similar to e.g. \citet{Shen:2011}. Furthermore, we have masked out regions in the near-IR which are strongly affected by telluric absorption.

For H$\alpha$ we first fit a local power-law continuum to wavelength regions free from emission lines. To the continuum subtracted spectrum we fit a line model over the range $6200-7000$\AA\, rest frame consisting of up to three Gaussians for the broad H$\alpha$. We add a set of narrow lines if justified by the data including a single Gaussian each for narrow H$\alpha$, [\ion{N}{2}] $\lambda,\lambda6548,6584$ and [\ion{S}{2}] $\lambda,\lambda6717,6731$, whose line widths and velocity offsets are tied together, while the flux ratio of the [\ion{N}{2}] lines is fixed to 2.96.

For H$\beta$ we fit a local pseudo-continuum, consisting of a power-law continuum and an optical iron template \citep{Boroson:1992}. In a few cases where the spectral quality in particular at the edges of the near-IR spectra does not allow a reliable fit of the pseudo-continuum, we instead fit a power-law continuum to areas free of emission lines and strong iron contribution and allow for iron contribution by modeling the strongest iron features around H$\beta$ and [\ion{O}{3}] by a double Gaussian centered at $\lambda\lambda4924,5018$\AA\, \citep[e.g.][]{Schulze:2009}. 
We fit a line model over the range $4700-5100$\AA\, to the pseudo-continuum subtracted spectra. We fit again up to three Gaussians for the broad H$\beta$ line. We use up to two Gaussians each to model the narrow  [\ion{O}{3}] $\lambda\lambda4959,5007$ lines which are coupled together in their shape and their line ratio of 3.0. This allows to capture the often asymmetric shape of the [\ion{O}{3}] line profile with a core and a blue wing component. The narrow H$\beta$ line is fitted with a single Gaussian with its velocity offset and line width fixed to the core component of [\ion{O}{3}].

We perform all our measurements (line and continuum luminosities, FWHM) from our best fit spectral model. Uncertainties on these parameters are derived via a Monte-Carlo approach. For each spectrum we generate 100 simulated spectra by adding  Gaussian random noise to the spectra, with the standard deviation at each pixel taken from the flux error. Each simulated spectrum is automatically fitted with the same model and the uncertainties for each measured parameter are obtained as the 1$\sigma$ dispersion from the fits to the set of simulated spectra.

We here use the FWHM of the broad component of the Balmer lines as the preferred measure of line width. Compared to the line dispersion, FWHM is less dependent on the wings of the profile and thus tends to be more robust at low signal-to-noise. Furthermore, FWHM is the reported width measure in the literature we use for our comparison with the non-BAL QSOs.

We have re-measured the systemic redshifts for the LoBAL QSO sample from our spectral models. In cases where [\ion{O}{3}] is significantly detected we base our redshift estimate on the model peak of the [\ion{O}{3}] $\lambda5007$ emission line, while we use the model peak of the total H$\alpha$ profile otherwise. Our near-IR redshift measurements are given in Table~\ref{tab:sample}, together with the optical SDSS redshift determined by \citet{Hewett:2010}.  Given the potential difficulty of obtaining a reliable redshift based on the BAL affected UV-lines, the \citet{Hewett:2010} redshifts are in good agreement with the near-IR redshifts, with only three objects having $|z_{\rm{HW}}-z_{\rm{NIR}}|>0.02$ (and all within 0.07). Excluding these outliers,  when defining the difference between the two redshift estimates by $c(z_{\rm{HW}}-z_{\rm{NIR}})/(1+z_{\rm{NIR}})$ we find a mean offset of $\sim150$~km s$^{-1}$ and a dispersion of $\sim820$~km s$^{-1}$.

The continuum and line measurements are given in Table~\ref{tab:prop}. Example line fits are shown in Fig.~\ref{fig:specex}. Line fits for the full sample are given in the Appendix. 

We note that for SDSS~J0841+2005 \citet{Stern:2017} recently published a $K$-band MOSFIRE/Keck spectrum with higher signal-to-noise than the one we present here. Our measurements of redshift, H$\alpha$ FWHM and luminosity are fully consistent with the values obtained from the MOSFIRE spectrum.

\begin{figure*}
\centering
\includegraphics[width=16cm,clip]{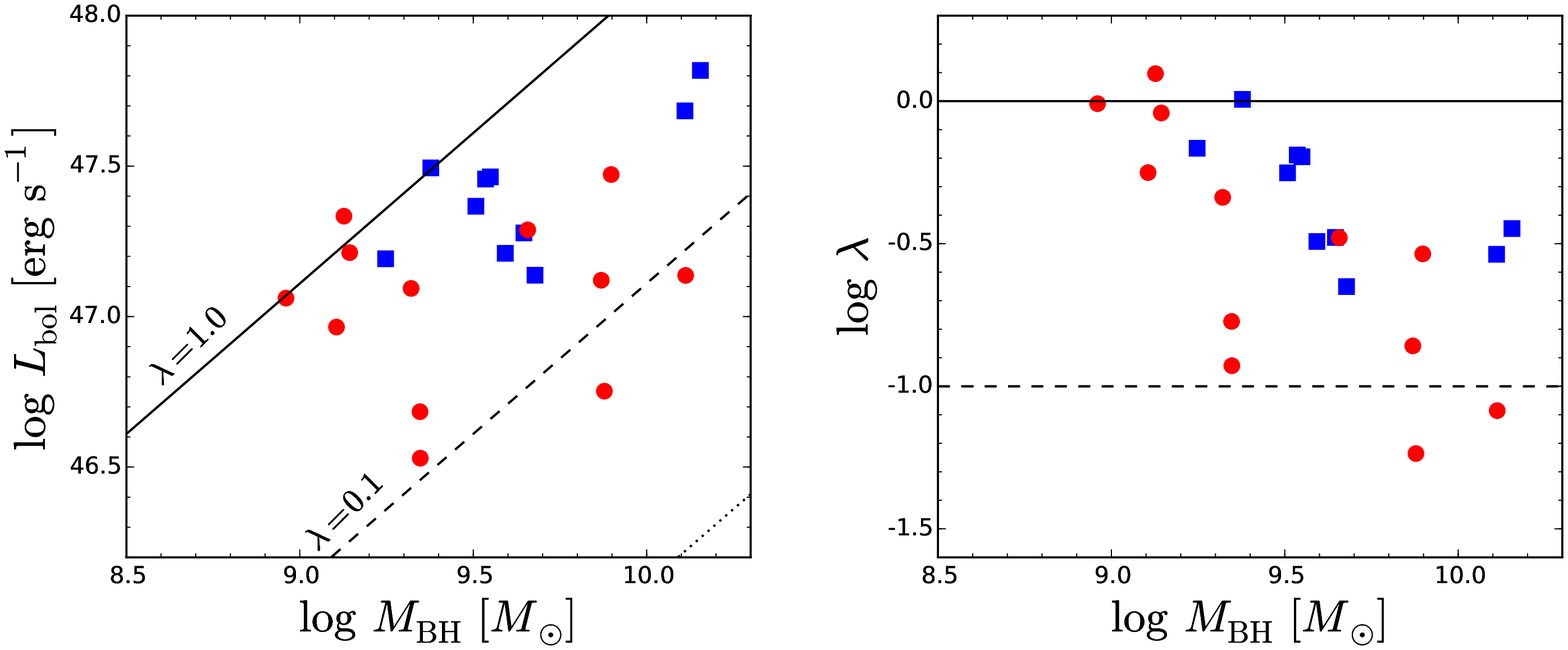}
\caption{Black hole mass - luminosity plane (left panel ) and black hole mass - Eddington ratio plane (right panel) for our LoBAL samples at $z\sim1.5$ (red circles) and $z\sim2.3$ (blue squares). The solid, dashed and dotted lines indicate Eddington ratios of one, 10\% and 1\% respectively.}
\label{fig:mlplane}
\end{figure*}

\subsection{Black hole masses and Eddington ratios}
We derive estimates of the black hole mass for our sample based on the virial method for single epoch broad line AGN spectra \citep[e.g.][]{McLure:2004, Greene:2005, Vestergaard:2006}. 
We have H$\alpha$ measurements for all objects in our sample and in addition H$\beta$ measurements for a major fraction (in particular for the $z\sim1.5$ sample). We will use black hole mass estimates based on the broad H$\alpha$ line as our primary black hole mass estimator and use H$\beta$ for comparison (see section~\ref{sec:linecomp}). 
While H$\beta$ is most directly calibrated to reverberation mapping results \citep{Vestergaard:2006}, H$\alpha$ serves equally as a reliable black hole mass estimator \citep{Greene:2005,Shen:2012,Mejia:2016} and for our sample has the advantage of higher signal-to-noise and being less affected by reddening. We therefore argue and show below that H$\alpha$ is indeed the prefered black hole mass estimator for our sample.
At the luminosities of our samples the host galaxy contribution to $L_{5100}$ is negligible \citep{Shen:2011}.

As our reference relation we use the formula for H$\beta$ by  \citet{Vestergaard:2006}:
\begin{equation}
\mbh (\rm{H}\beta)= 10^{6.91} \left( \frac{L_{5100}}{10^{44}\,\mathrm{erg\,s}^{-1}}\right)^{0.50} \left( \frac{\mathrm{FWHM}}{3000\,\mathrm{km\,s}^{-1} }\right)^2 M_\odot  \label{eq:mbhHb}
\end{equation} 

This relationship is directly derived from reverberation mapping studies. We use it to compute black hole masses from H$\beta$ and calibrate our H$\alpha$ black hole masses to this relation. For H$\alpha$ we use the FWHM and luminosity of the broad H$\alpha$ line to estimate the black hole mass as done in previous work \citep{Greene:2005,Shen:2012,Jun:2015}.
We use the relations between H$\alpha$ and H$\beta$ FWHM from \citet[][their Equation~4]{Jun:2015}, derived from a compilation of studies covering a large luminosity range. For the scaling relation between $L_{5100}$ and $L_{\rm{H}\alpha}$ we adopt the scaling presented in the same paper (their Equation 2). This relation is consistent with \citet{Greene:2005} over the lower luminosity range covered in that study, but at the same time provides a better fit to the high luminosity regime, as studied in \citet{Shen:2012}, \citet{Jun:2015} and also in this work.

Combining these relations gives the following virial black hole mass estimator for H$\alpha$:
\begin{equation}
\mbh (\rm{H}\alpha)= 10^{6.711} \left( \frac{L_{\rm{H}\alpha}}{10^{42}\,\mathrm{erg\,s}^{-1}}\right)^{0.48} \left( \frac{\mathrm{FWHM}}{3000\,\mathrm{km\,s}^{-1} }\right)^{2.12} M_\odot.  \label{eq:mbhHa}
\end{equation} 

To compute bolometric luminosities we use the broad H$\alpha$ luminosity $L_{\rm{H}\alpha}$ adopting a bolometric correction factor of 130 \citep{Stern:2012}, i.e. $L_{\rm{bol}}=130 L_{\mathrm{H}\alpha}$. 
The Eddington ratio is then given by $\er=L_{\rm{bol}}/L_{\rm{Edd}}$, where $L_{\rm{Edd}}$ is the Eddington luminosity for the object, given its black hole mass.

Since LoBAL QSOs typically show high levels of dust extinction (including our sample as shown in Sections~\ref{sec:sed} and \ref{sec:stack}) this might also affect our measurement of $L_{\rm{H}\alpha}$ and thus \mbh and $L_{\rm{bol}}$. Assuming a typical value of $E(B-V)=0.14$  would indicate an underestimate of the flux at 6565\AA{} of 0.12~dex, assuming an SMC-like extinction curve. To test for a systematic bias in $L_{\rm{H}\alpha}$ in our sample, we use an alternative estimate of the intrinsic luminosity, based on the mid-IR luminosity at $4.6\mu$m, which for typical LoBAL $E(B-V)$ values is largely unaffected by dust extinction. This luminosity is obtained from the Wide-Field Infrared Survey Explorer (\textit{WISE}) $W2$ magnitude (see Section~\ref{sec:sed} for details on the mid-IR data for our sample). We use the quasar spectral energy distribution template from \citet{Richards:2006} to obtain the typical ratio to rest-frame 5100\AA{} and convert the $W2$ magnitude to $L_{5100}$ (the typical ratio for our sample is $\sim1$). We then use the relation by \citet{Jun:2015} to convert to $L_{\rm{H}\alpha}$. Comparing this intrinsic $L_{\rm{H}\alpha}$ estimate with our measurements for the $z\sim1.5$ sample we find zero offset with a dispersion of 0.28~dex. The $z\sim2.3$ sample on average even shows larger measured $L_{\rm{H}\alpha}$ than the intrinsic estimate. We thus conclude that we do not see evidence for a systematic bias in $L_{\rm{H}\alpha}$ compared to the general quasar population due to dust extinction.

In Fig.~\ref{fig:mlplane} we show the location of our two LoBAL samples ($z\sim1.5$ and $z\sim2.3$) in the black hole mass - luminosity plane and the black hole mass - Eddington ratio plane. We can already see that given the high luminosity limit of the respective samples they cover a broad range of \mbh and \er with $8.7<\log\, \mbh<10.2$ and $-1.1<\log\, \er<0.3$. As expected by our selection, the $z\sim2.3$ sample shows on average higher bolometric luminosities and also more massive black holes. We will discuss the black hole masses and Eddington ratios of these LoBAL QSOs in comparison to normal QSOs in more detail in section~\ref{sec:eddcomp}.

\begin{figure*}
\centering
\includegraphics[width=18cm,clip]{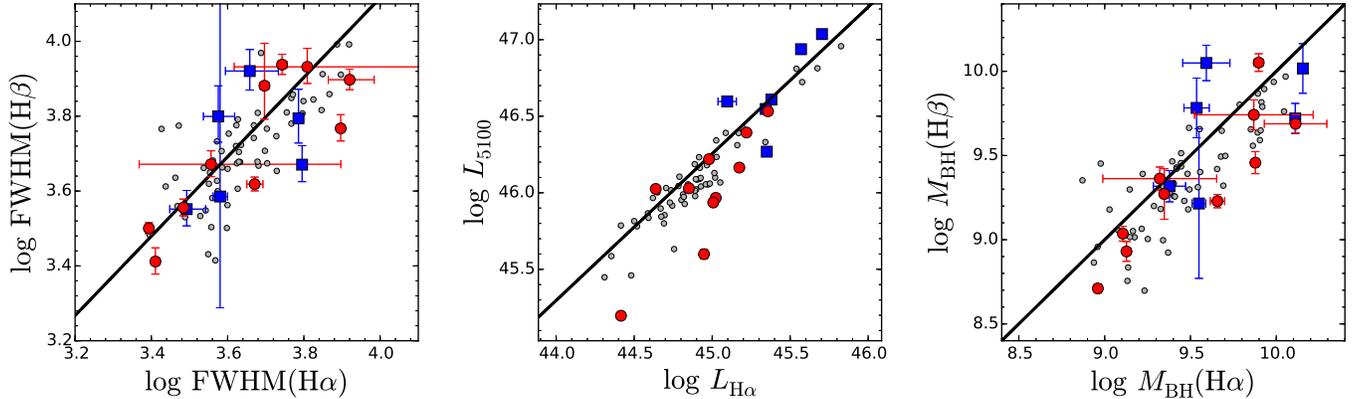}
\caption{Comparison of  spectral measurements and derived black hole masses between H$\alpha$ and H$\beta$ for the LoBAL QSOs at $z\sim1.5$ (red circles) and at $z\sim2.3$ (blue squares) and for a representative non-BAL QSO sample at the same redshifts and luminosities from \citet{Shen:2012} (gray circles). \textit{Left panel:} Comparison of H$\alpha$ and H$\beta$ FWHMs. The solid black line shows their relation from \citet{Jun:2015}. \textit{Middle panel:} Comparison between $L_{\rm{H}\alpha}$ vs. $L_{5100}$, with the relation from \citet{Jun:2015} shown by the solid black line. 
\textit{Right panel:} Comparison of the derived estimates of \mbh. The solid black line shows the one-to-one relation.}
\label{fig:linecompare}
\end{figure*}

\subsection{Line comparison} \label{sec:linecomp}
The primary goal of this paper is not to address the cross-calibration of different black hole mass estimators based on different broad lines. While this is a highly relevant topic which has drawn significant attention \citep{McGill:2008,Assef:2011,Trakhtenbrot:2012,Shen:2012,Ho:2012,Park:2013,Matsuoka:2013,Jun:2015,Mejia:2016,Bisogni:2017}, our sample is by design not particularly suited for such studies, since the broad \ion{Mg}{2} and \ion{C}{4} lines are severely affected by the broad absorption throughs and therefore do not  serve as black hole mass estimator. This is in general not the case for the Balmer lines, apart from the case of SDSS~J1019$+$0225 which shows unusual blue shifted absorption features in both H$\alpha$ and H$\beta$ (see Fig.~\ref{fig:allspec_z15}; for the emission line fit of this object we have masked out this region).

We here investigate the consistency of the broad line widths, luminosities and black hole mass estimates between H$\alpha$ and H$\beta$ for our LoBAL quasar sample in comparison to the general quasar population. In Fig.~\ref{fig:linecompare} we compare the FWHMs,  $L_{\rm{H}\alpha}$ vs. $L_{5100}$ and the resulting black hole mass estimates. The correlation between H$\alpha$ and H$\beta$ FWHM from \citet{Jun:2015}, the relation between $L_{\rm{H}\alpha}$ and $L_{5100}$ from \citet{Jun:2015} as well as the one-to-one relation for \mbh are indicated by the solid black lines in each of the panels. In addition we show the values for a representative non-BAL QSO sample of 60 objects at similar redshift and luminosity from \citet{Shen:2012} underlying as gray circles. They targeted normal luminous ($L_{\rm{bol}}>$ a few $\times 10^{46}$~erg s$^{-1}$) quasars at redshift $1.5<z<2.2$, with the majority of them at $1.5<z<1.7$.

We find a reasonably good correlation for the FWHM, with the $z\sim1.5$ sample showing a mean offset of $-0.04$ and a standard deviation of 0.10~dex. The $z\sim2.3$ sample shows a slightly larger scatter ($\sigma=0.14$~dex), due to the typical lower signal-to-noise in the H$\beta$ line, but shows basically no mean offset (mean $\Delta \log\,$FWHM$=0.001$).

For the luminosity comparison we find the $z\sim2.3$ sample to be in good agreement with the relation for normal quasars. For the $z\sim1.5$ sample about half of the sample follows the relation for normal quasars by \citet{Jun:2015}, while the other half shows lower $L_{5100}$ than expected.
This lower  $L_{5100}$ is  caused by increased intrinsic reddening in these objects. Indeed we find these objects to show overall the strongest reddening based on the spectral energy distribution and the UV-optical spectral shape, as discussed further below and shown in Figs.~\ref{fig:allspec_z15}-\ref{fig:allspec_z23b}.  We conclude that for our LoBAL sample $L_{\rm{H}\alpha}$ is the spectroscopic intrinsic luminosity indicator which is least affected by dust reddening and therefore our preferred estimator of $L_{\rm{bol}}$ and \mbh.

The right panel in Fig.~\ref{fig:linecompare} compares the derived black hole masses using Equations~\ref{eq:mbhHb} and \ref{eq:mbhHa}. We find a clear correlation between both mass estimates. For the $z\sim1.5$ sample  both \mbh estimates are in good agreement. 
We find a mean difference of 0.18 and a standard deviation $\sigma=0.19$~dex. The offset towards less massive H$\beta$ \mbh  is at least partly due to the objects with lower $L_{5100}$ discussed above.
The $z\sim2.3$ sample again shows larger scatter due to the typical low S/N for H$\beta$ in this sample but the black hole masses are overall consistent (mean $\Delta \log  \mbh=-0.01$ and $\sigma=0.30$~dex). This verifies the reliability of using H$\alpha$ as a black hole mass estimator \citep[e.g.][]{Greene:2005,Ho:2012,Mejia:2016}. We conclude that H$\alpha$ remains a reliable black hole mass estimator also for the quasar sub-population of LoBAL QSOs. We therefore use \mbh and \er based on broad H$\alpha$ for the rest of the paper as listed in Table~\ref{tab:prop}.

\begin{figure*}
\centering
\includegraphics[width=18cm,clip]{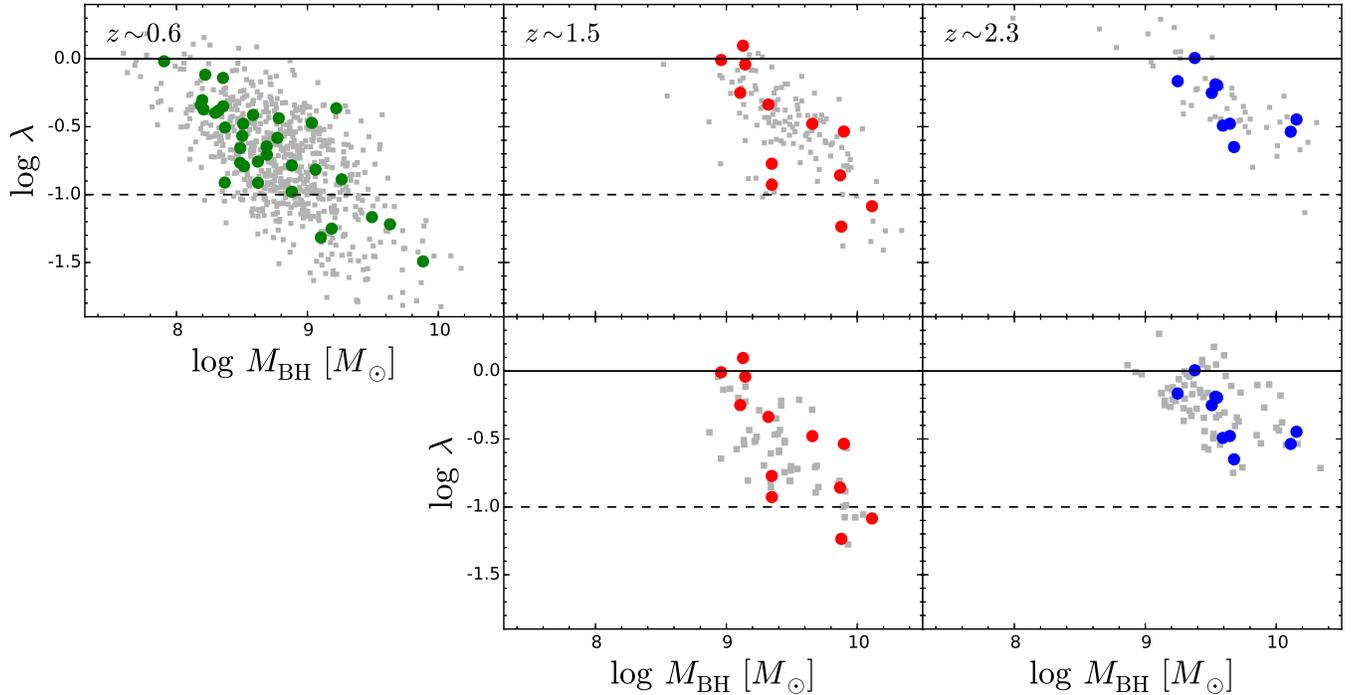}
\caption{Comparison of the black hole mass and Eddington ratio distribution of the LoBAL QSOs (green/red/blue circles) to a matched non-BAL QSO sample (gray squares). In the upper row the matched sample is taken from the SDSS DR7 quasar catalog, matched in $z$ and 2MASS magnitude. In the lower row we show the match to near-IR spectroscopy samples from the literature of similar $z$ and luminosity as our sample (see text for details). The solid and dashed lines indicate Eddington ratios of one and 10\% respectively.}
\label{fig:mecompare}
\end{figure*}

\section{Discussion}  \label{sec:discussion}
\subsection{Are LoBAL QSOs in an Eddington limited accretion phase?}  \label{sec:eddcomp}
A potential implication of the scenario that LoBAL QSOs represent a special evolutionary phase, corresponding to a young AGN, is that they might on average accrete at a higher rate than non-BAL QSOs, since they should still have ample fuel supply while just being in the process of blowing off their dusty envelope. Thus we would expect to find on average higher Eddington ratios for our LoBAL quasar sample compared to a luminosity matched non-BAL quasar sample. This is the hypothesis we test in this section.  

\begin{deluxetable*}{l  ccc  ccc  ccc}
\tabletypesize{\scriptsize}
\tablecaption{Comparison of \mbh and \er distributions of LoBALs and non-BALs.}
\tablewidth{18cm}
\tablehead{  & \multicolumn{3}{c}{$\log \mbh$} & \multicolumn{3}{c}{$\log \er$} & \multicolumn{3}{c}{$\log L_{\rm{bol}}$} \\ \noalign{\smallskip}
\colhead{Sample} & \colhead{LoBAL} & \colhead{non-BAL} & \colhead{$p_{\rm{KS}}$} & \colhead{LoBAL} & \colhead{non-BAL} & \colhead{$p_{\rm{KS}}$} & \colhead{LoBAL} & \colhead{non-BAL} & \colhead{$p_{\rm{KS}}$}
}
\startdata
$z\sim0.6$ - SDSS &   8.70 $\pm$  0.08 &  8.83 $\pm$  0.02 & 0.0233 & -0.65 $\pm$  0.06 & -0.76 $\pm$  0.02 & 0.224 & 46.15 $\pm$  0.05 & 46.17 $\pm$  0.01 & 0.535 \\ 
$z\sim1.5$ - SDSS &   9.48 $\pm$  0.11 &  9.50 $\pm$  0.03 & 0.488 & -0.54 $\pm$  0.12 & -0.51 $\pm$  0.03 & 0.534 & 47.05 $\pm$  0.08 & 47.09 $\pm$  0.02 & 0.951 \\ 
$z\sim2.3$ - SDSS &   9.64 $\pm$  0.09 &  9.55 $\pm$  0.06 & 0.761 & -0.34 $\pm$  0.06 & -0.25 $\pm$  0.05 & 0.558 & 47.41 $\pm$  0.07 & 47.40 $\pm$  0.04 & 0.925 \\ 
\noalign{\smallskip} \hline \noalign{\smallskip}
$z\sim1.5$ - NIR &  9.48 $\pm$  0.11 &  9.44 $\pm$  0.04 & 0.838 & -0.54 $\pm$  0.12 & -0.58 $\pm$  0.04 & 0.604 & 47.05 $\pm$  0.08 & 46.97 $\pm$  0.04 & 0.125 \\
$z\sim2.3$ - NIR & 9.64 $\pm$  0.09 &  9.53 $\pm$  0.05 & 0.466 & -0.34 $\pm$  0.06 & -0.12 $\pm$  0.03 & 0.0392 & 47.41 $\pm$  0.07 & 47.52 $\pm$  0.04 & 0.353 \\
\enddata
\tablecomments{We list the mean \mbh, \er and  $L_{\rm{bol}}$ for the LoBAL sample and respective comparison sample, together with it's uncertainty and the K-S test probability.}
\label{tab:test}
\end{deluxetable*}

To construct our non-BAL comparison sample we follow two different approaches. The first uses the SDSS DR7 quasar catalog \citep{Shen:2011} to find a large sample matched directly to our LoBAL sample in redshift and either $H$-band magnitude (for the $z\sim1.5$ sample) or $K$-band magnitude (for the $z\sim2.3$ sample). We then use black hole masses derived from \ion{Mg}{2} (for the  $z\sim1.5$ sample) or \ion{C}{4} (for the $z\sim2.3$ sample) to compare the observed black hole mass and Eddington ratio distributions to our LoBAL sample. This has the advantage of having a large comparison sample with the closest match to the LoBAL quasar sample, but at the risk of introducing potential systematics due to the use of different black hole mass estimators.

We therefore also use representative near-IR spectroscopic samples of non-BAL quasars with measured properties of either broad H$\alpha$ or H$\beta$ which cover the same redshift and near-IR magnitude range as our samples. For the $z\sim1.5$ sample we use the study by \citet{Shen:2012}, restricted to $1.5<z<1.8$, containing 55 quasars with H$\alpha$ measurements.  They obtained near-IR observations of normal SDSS quasars selected for high S/N optical spectra, which lead to bolometric luminosities of their sample of $\log\, L_{\rm{bol}}>46.4$, well matched to our LoBAL sample.

For the $z\sim2.3$ sample we construct a comparison sample based on the study by \citet{Coatman:2017}. They presented a large sample of 230 luminous quasars at redshift $1.5<z<4.0$ with either broad H$\alpha$ and/or H$\beta$ measurements from near-IR spectroscopy. We restrict their sample to objects within $2.0<z<2.6$ with broad H$\alpha$ measurements and $L_{\rm{bol}}>10^{47}$~erg s$^{-1}$ to approximately match their sample to our LoBAL sample. The luminosity cut corresponds to the flux limit within our LoBAL sample. This selection results in a $z\sim2.3$  comparison sample of 70 non-BAL QSOs. For both the $z\sim1.5$ and $z\sim2.3$  comparison sample we estimated black hole masses and bolometric luminosities in the same way as for our LoBAL sample from the reported FWHMs and luminosities.

In addition, we augment our two redshift bins with a lower $z$-bin again based on the BAL catalog from \citet{Allen:2011}. We select all objects with BI(\ion{Mg}{2})$>0$ within the redshift range $0.4<z<0.9$, allowing simultaneous coverage of \ion{Mg}{2} and H$\beta$ in the optical SDSS spectra. We further require detection by 2MASS in $J$-band and a black hole mass estimate based on H$\beta$ in the SDSS DR7 quasar catalog from \citet{Shen:2011}. This gives a $z\sim0.6$ LoBAL quasar sample of 34 objects. We match it with a non-BAL quasar sample (BI(\ion{Mg}{2})$=0$), where we match them as close as possible in redshift and $J$-band magnitude with 20 non-BALs for every LoBAL.

 \begin{figure*}
\centering
\includegraphics[width=17cm,clip]{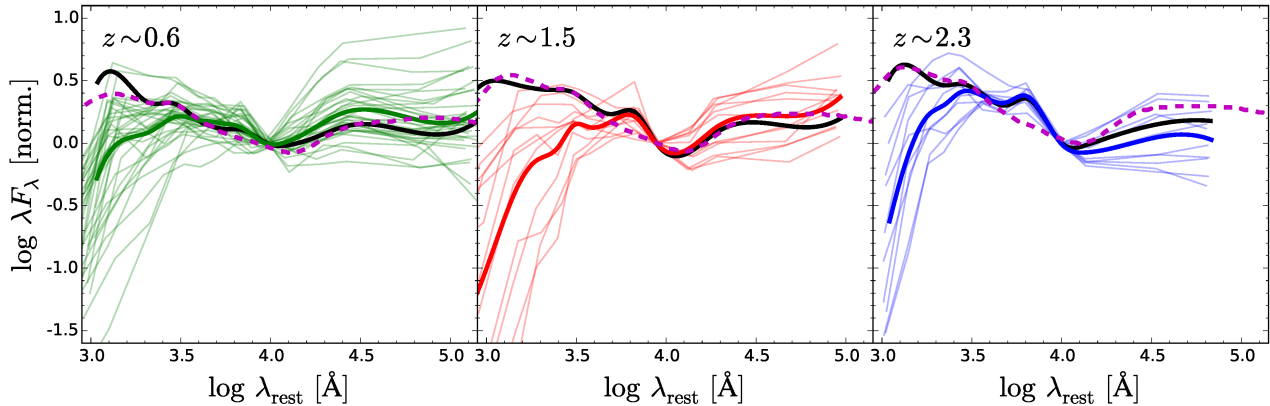} 
\caption{The spectral energy distributions (SEDs) for the individual LoBAL QSOs (green: $z\sim0.6$, red: $z\sim1.5$, blue $z\sim2.3$) are shown as thin lines, while their geometric mean SED is shown by the thick line. We compare this LoBAL SED with the geometric mean SED for the matched non-BAL samples (thick black line) and the SED template by \citet{Richards:2006} (dashed magenta line). For the latter we use their SED template for all QSOs at $z\sim1.5$ and $z\sim2.3$ and their optically faint ($\log L<46.02$) SED template at $z\sim0.6$ to approximately match the luminosity range of these samples.}
\label{fig:sed} \vspace{0.5cm}
\end{figure*}

We show the comparison between these samples in the \mbh-\er-plane in Fig.~\ref{fig:mecompare}. We see no apparent difference in the  \mbh or \er distribution of LoBAL QSOs and non-BAL QSOs. To quantify this visual impression we performed a 2-sample Kolmogorov-Smirnov (K-S) test as well as an Anderson-Darling (A-D) test on the distributions of $\log L_{\rm{bol}}$, \mbh and \er for each of the above 5 combinations of LoBAL sample and matched sample to test the null-hypothesis that the LoBAL QSO and the matched non-BAL QSO sample are drawn from the same distribution. The results for the K-S test as well as the mean values of the distributions are given in Table \ref{tab:test}. The A-D test provided consistent results. In no case do we find a statistically significant difference between any of these distributions, i.e. the distributions are statistically indistinguishable between the LoBAL and non-BAL QSO samples. In particular we do not see clear evidence for a higher Eddington ratio in the LoBAL samples. The $z\sim0.6$ sample shows slightly higher mean \er and lower \mbh by $\sim0.1$~dex for the LoBAL QSO sample, qualitatively consistent with \citet{Zhang:2010}, but not at a high significance for the sample we use here. More important, at $z>1$ we cannot confirm such a trend.
Our results are robust against the details of the matching and of the black hole mass estimator in use. Potential systematics between different lines for estimating \mbh would tend to increase the difference in the apparent distributions of \mbh and \er.

We conclude that we do not find evidence for LoBAL QSOs constituting a separate population in terms of their \mbh and \er. They are rather consistent with having the same black hole mass and Eddington ratio distributions as non-BAL QSOs.

We have also tested for any correlation of \mbh, \er and $\log L_{\rm{bol}}$ with the BAL properties, as measured by \citet{Allen:2011}. We do not find any significant correlation of these with their balnicity BI, mean BAL depth or the minimum and maximum velocities, based on their Spearman rank-order correlation coefficients.

\subsection{LoBAL Spectral Energy Distribution} \label{sec:sed}
We next test if our LoBAL sample shows any significant difference compared to the matched non-BAL sample in their Spectral Energy Distribution (SED). All of our targets posses multi-wavelength photometry from the far-UV to the mid-IR. In particular, we collect data from four surveys which provide 14 bands in total.  Optical data is taken from the SDSS DR7 quasar catalog \citep{Schneider:2010} in the $u, g, r, i$ and $z$ bands. The near-IR data in the $J$, $H$ and $K$ bands comes from the 2MASS catalog \citep{Skrutskie:2006}, with the matching provided by \citet{Schneider:2010}. We augment this photometry with UV data obtained by the all-sky Galaxy Evolution Explorer (\textit{GALEX}) space mission \citep{Martin:2005}. \textit{GALEX} provides measurements in the far-UV (FUV: 1350 to 1750 \AA{} ) and  the near-UV (NUV: 1750 to 2750 \AA{}). Finally, we add mid-IR (MIR) data from the all-sky Wide-Field Infrared Survey Explorer \citep[\textit{WISE}; ][]{Wright:2010} mission, covering four bands at $3.4, 4.6, 12$ and $22 \mu$m respectively. We take the \textit{WISE} data for our SDSS QSOs from  \citet{Lang:2016}, obtained from forced photometry of the WISE all-sky release imaging at SDSS positions. All of the  photometric data have been corrected for Galactic extinction \citep{Schlegel:1998}. We correct for missing data in a similar way to \citet{Richards:2006}. We use the AGN SED template by \citet{Richards:2006} normalized at the neighboring band to derive the typical flux density in the missing band. For the LoBAL QSOs we add a typical LoBAL reddening to the SED template.

In Fig.~\ref{fig:sed} we present the SED for each individual source from our LoBAL QSO sample as well as their geometric mean SED in the three redshift bins studied above. We note that a detailed SED modeling for our sources is beyond the scope of this work, we here highlight the average SED shape in comparison to the general quasar population. For this, we compare the LoBAL photometry with the SED obtained in the same way from the matched SDSS non-BAL QSO sample presented in Section~\ref{sec:eddcomp} (black solid line) as well as with the AGN SED template from \citet{Richards:2006} (magenta dashed line). We do not correct the SED for contributions other than the AGN continuum, like host galaxy contamination or the contribution by emission lines. With the matching of the LoBAL and the non-BAL samples these will contribute in a similar way to the SED and we are here not interested to determine the intrinsic SED of LoBAL QSOs but only in the comparison to the general quasar population. 
The enhancement in the band around 6400\AA{} we see in the $z\sim1.5$ and $z\sim2.3$ samples in comparison to the \citet{Richards:2006} SED template is likely due to the contribution from the H$\alpha$ line (given its large equivalent width of $\sim400$~\AA{}), which is by design of these samples centered on the respective band.

 \begin{figure}
\centering
\resizebox{\hsize}{!}{ \includegraphics[clip]{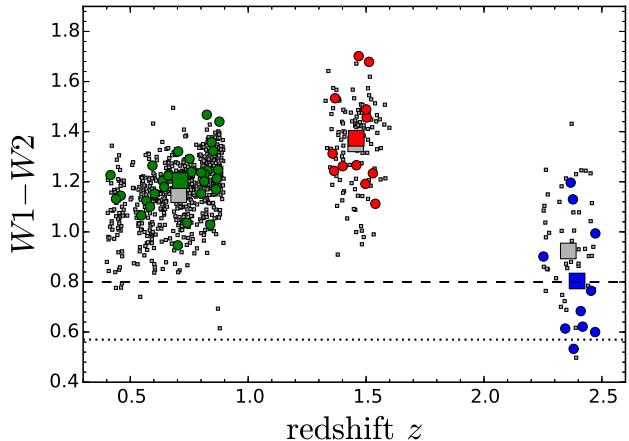} }
\caption{\textit{WISE} $W1-W2$ colors for our LoBAL QSO samples at $z\sim0.6$ (green), $z\sim1.5$ (red) and $z\sim2.3$ (blue) and the matched non-BAL QSO samples (grey). The large squares show the mean values for the respective samples. The dashed black line indicates the AGN \textit{WISE} selection criteria $W1-W2\geq0.8$ proposed by \citet{Stern:2012b}, while the dotted black line indicates $W1-W2\geq0.57$ proposed by \citet{Wu:2012}.}
\label{fig:wise} 
\end{figure}

Comparing the thus constructed SEDs, the most obvious difference between our LoBAL QSOs and non-BAL QSOs is the significantly reduced flux at $\lambda_{\rm{rest}} \lesssim 3000$\AA{} leading to red colors, in agreement with previous work \citep{Weymann:1991,Reichard:2003, Gibson:2009}. The redder color is usually interpreted as excess dust reddening in LoBAL quasars \citep[e.g.][]{Sprayberry:1992}.

Apart from the stronger reddening in the rest-frame UV we do not see a clear difference between our LoBAL samples and the non-BAL QSOs. In particular there is no apparent difference on the red side of the accretion disk emission at $\lambda_{\rm{rest}} \gtrsim 5000$\AA{} and in the dust torus emission in the near-to-mid IR. 

The mid-IR emission in QSOs originates from reprocessed UV-optical emission from the so-called torus, a cold, dusty obscuring medium distributed on spatial scales of $>1$~pc. The evolutionary scenario for LoBAL QSOs implies a large dust covering fraction of the BAL wind, i.e. the BAL is visible along most orientation angles but only present in a small fraction of the quasar population. As pointed out by \citet{Gallagher:2007}, in this case it might be expected that BAL QSOs will have enhanced mid-IR emission due to the larger emitting volume of dust. We do not find evidence for such an enhancement and thus no support for the evolutionary scenario, consistent with previous work on HiBAL and LoBAL QSOs \citep[][but see \citet{DiPompeo:2013} for a different result for radio-loud BAL QSOs]{Gallagher:2007,Lazarova:2012}.

In Fig.~\ref{fig:wise} we show the \textit{WISE} $W1-W2$ colors for our LoBAL QSO samples and the matched non-BAL QSO samples. Both populations are consistent with being drawn from the same population, based on a 2-sample Kolmogorov-Smirnov test. All LoBALs in the two samples at $z<2$ satisfy the \textit{WISE} AGN selection criteria from \citet{Stern:2012b} $W1-W2\geq0.8$. At $z>2$ the criteria by  \citet{Stern:2012b} is less complete, as also shown for our $z\sim2.3$ LoBAL sample. \citet{Wu:2012} proposed a less strict criteria $W1-W2\geq0.57$ to select $z<3.2$ quasars. Most of our LoBALs, also those at $z>2$, satisfy this criteria.
While optical color selection is biased against LoBAL QSOs due to their red colors, \textit{WISE} MIR selection is a promising technique to obtain an unbiased census of the luminous LoBAL population at $z\lesssim2$ and possibly beyond.

We conclude that apart from the well known higher reddening of LoBAL QSOs in the UV, their optical ($\lambda_{\rm{rest}} \gtrsim 4000$\AA{}) to MIR SEDs are consistent with the general quasar population.
  
 \begin{figure}
\centering
\resizebox{\hsize}{!}{\includegraphics[clip]{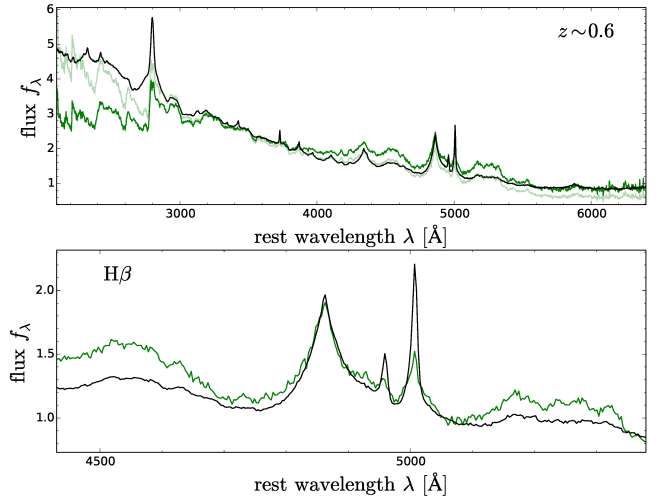}}
\caption{Comparison of the geometric mean composite spectrum for the LoBAL sample at $z\sim0.6$ (green) with that of a matched non-BAL sample from SDSS DR7 (black). The upper panel shows the full spectrum from 2100-6400\AA{}. A reddening corrected LoBAL composite is shown there by the light green line, assuming $E(B-V)=0.14$ and an SMC-like extinction curve.The lower panel shows a zoom-in on the H$\beta$ region, where we normalized the spectra at 5100\AA{} and show only the extinction corrected LoBAL composite.}
\label{fig:stack_z06} 
\end{figure}
  
\begin{figure*}
\centering
\includegraphics[width=18cm,clip]{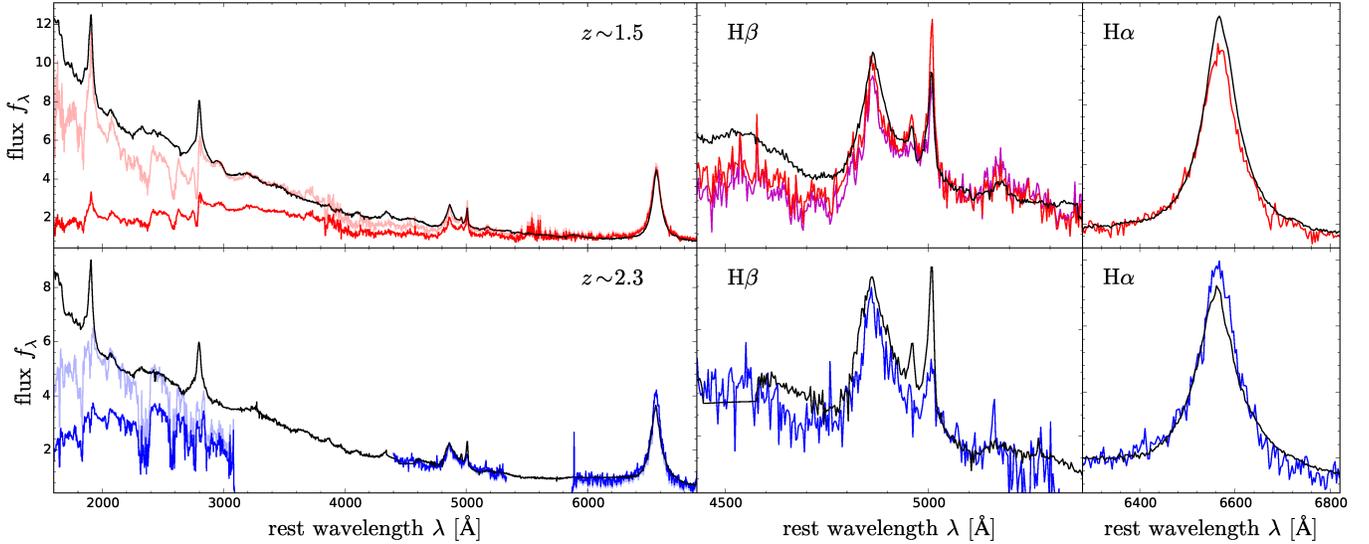}
\caption{Comparison of the geometric mean composite spectra for the LoBAL samples at $z\sim1.5$ (upper panels) and $z\sim2.3$ (lower panels) with  those of a representative non-BAL sample from \citet{Shen:2016} (black). The left panels shows the full spectrum from 1600-6900\AA{}. A reddening corrected LoBAL composite is shown by the light red (blue) line, assuming $E(B-V)=0.17 (0.1)$. The center and right panels shows a zoom-in on the H$\beta$ and H$\alpha$ regions, with the spectra normalized at 5100\AA{} and 6350\AA{} respectively. Again we only show the extinction corrected LoBAL composite spectra. The magenta spectrum for H$\beta$ at $z\sim1.5$ shows the composite when excluding two cases of strong Balmer absorption.}
\label{fig:stack_nir} \vspace{0.3cm}
\end{figure*}

\subsection{Stacked spectra}  \label{sec:stack}
To investigate potential differences in the spectral continuum and emission line properties between our LoBAL QSO sample and the general QSO population, we generate composite spectra for both populations. We generate individual composite spectra for the three LoBAL samples at $z\sim0.6$ (based on SDSS DR7 spectra), $z\sim1.5$ and $z\sim2.3$. As non-BAL comparison sample for the $z\sim0.6$ sample we use the matched SDSS QSO sample discussed in Section~\ref{sec:eddcomp} including 680 objects. For the $z\sim1.5$ and $z\sim2.3$ sample we use the sample from \citet{Shen:2012} with spectra made publicly available by \citet{Shen:2016}, restricted to the same broad redshift bin, including 55 and 5 objects respectively. The non-BAL samples cover the same luminosity range as their respective LoBAL QSO samples, so we are not affected by luminosity dependent effects like the Baldwin effect in lines like [\ion{O}{3}] \citep{Netzer:2004,Stern:2013,Zhang:2013} or the amount of host galaxy contribution for each $z$-bin comparison.

Each spectrum is shifted into rest-frame, re-binned to a common wavelength scale and normalized at 5100\AA{} for the $z\sim0.6$ sample and at 6350\AA{} otherwise and a stacked spectrum is generated using the geometric mean. Uncertainties are derived from bootstrapping the sample where we applied the Monte-Carlo approach discussed in \ref{sec:fitting} to every bootstrapped object.
The derived composite spectra for the three LoBAL samples and non-BAL QSO comparison samples are shown in Fig.~\ref{fig:stack_z06} and Fig.~\ref{fig:stack_nir}. For the $z\sim2.3$ LoBAL sample we use BOSS spectra for the rest-frame UV coverage when available (in 9/10 cases), based on SDSS DR12 \citep{Alam:2015}, applying the improved spectrophotometric calibration by \citet{Margala:2016}.

The most prominent continuum difference is again the significant reddening for the LoBALs in all three $z$-bins at $\lambda_{\rm{rest}} \lesssim 4000$\AA{}, consistent with previous LoBAL composite spectra studies \citep{Weymann:1991,Brotherton:2001,Reichard:2003,Zhang:2010}. In addition, the strong broad absorption throughs are clearly visible. The $z\sim2.3$ LoBAL sample shows the most extreme BAL properties as also seen in their individual spectra in Figs.~\ref{fig:allspec_z23} and \ref{fig:allspec_z23b}.
Assuming SMC-like dust extinction, our data are consistent with a reddening of $E(B-V)\sim 0.14, 0.17$ and $0.10$ for the $z\sim0.6$, $z\sim1.5$ and $z\sim2.3$ sample respectively. 

The rest-frame optical properties between the LoBAL QSOs and the non-BAL QSOs are at first sight remarkably similar. We do not see any major differences in the broad Balmer line profiles. \citet{Boroson:1992b} reported an excess  blue wing component in H$\alpha$ for their small LoBAL QSO sample compared to their control sample. We cannot confirm this trend for our sample.
The $z\sim0.6$ sample shows on average weak [\ion{O}{3}] and strong iron-emission, consistent with previous work on low-$z$ LoBAL QSOs \citep{Weymann:1991,Boroson:1992b,Zhang:2010} and the small sample of LoBALs (mainly) located at  $0.6<z<1.2$ from \citet{Runnoe:2013}. For our two intermediate $z$ samples these trends are less clear. Both the $z\sim1.5$ and the $z\sim2.3$ sample do not show significantly enhanced iron-emission. The $z\sim1.5$ LoBAL composite may even have \textit{weaker} optical \ion{Fe}{2} emission than the non-BAL composite.
The $z\sim2.3$ sample shows on average weak or absent [\ion{O}{3}] emission, however the  [\ion{O}{3}] observations for this sample suffer from small number statistics and low-signal-to-noise. Only one of six objects shows a clear strong [\ion{O}{3}] line.
The $z\sim1.5$ sample shows [\ion{O}{3}] emission at least as strong as for the non-BAL comparison sample. While the composite spectrum derived from all 12 $z\sim1.5$ LoBAL QSOs even shows enhanced [\ion{O}{3}], this is mainly driven by the two objects with broad Balmer absorption lines and very strong  [\ion{O}{3}] emission (Schulze et al., in prep). When excluding these two from the stack the  [\ion{O}{3}] profile of the LoBAL QSO composite is fully consistent with the non-BAL QSO composite (see magenta line in Fig.~\ref{fig:stack_nir}).

The [\ion{O}{3}] emission arises at larger distances from the nucleus in the Narrow Line Region (NLR) and is therefore largely an isotropic quantity. Significantly different [\ion{O}{3}] equivalent widths for LoBAL QSOs would be difficult to explain in a pure orientation scenario and rather support a large covering fraction of the BAL wind which might shield the NLR from part of the ionizing radiation. Unfortunately, our results are inconclusive on this. While we see reduced [\ion{O}{3}] in the $z\sim0.6$ and $z\sim2.3$ sample, the [\ion{O}{3}]  strength in the $z\sim1.5$ sample is consistent with the general quasar population.

\begin{deluxetable}{l  ccc}
\tabletypesize{\scriptsize}
\tablecaption{[\ion{O}{3}] FWHM in the composite spectra}
\tablewidth{8cm}
\tablehead{ \colhead{Sample} & \colhead{LoBAL} & \colhead{LoBAL$-$BA} & \colhead{non-BAL} 
}
\startdata
$z\sim0.6$ & $588\pm70$ &  & $634\pm6$ \\ 
$z\sim1.5$ & $914\pm96$  & $1036\pm156$ & $964\pm21$\\
$z\sim2.3$ & $1212\pm460$ & & $944\pm40$ \\
\enddata
\tablecomments{FWHM given in km s$^{-1}$, for the LoBAL composite and for the non-BAL composite. In column LoBAL$-$BA we show the result when excluding two cases of strong Balmer absorption in the composite.}
\label{tab:o3width}
\end{deluxetable}

Furthermore, the [\ion{O}{3}] profile in AGN often shows a broad blue wing component indicative of outflows on NLR scales. Connecting the  BAL wind originating on small scales and the large scale ionized outflows traced via [\ion{O}{3}] can help to understand the outflow phenomena and the role AGN winds play for AGN feedback \citep[e.g.][]{Fiore:2017}. If LoBAL QSOs are young AGN in the process of blowing off their dusty environment, an ubiquitous existence of powerful outflows might be expected.
Indeed, our LoBAL sample shows several cases of broad [\ion{O}{3}] lines (FWHM$>1000$ km s$^{-1}$) and extended wings indicating the presence of powerful outflows in these objects. But while their demographics are still not well understood,  signatures of outflows seem to be common in the general luminous quasar population, in particular at high-$z$  \citep{Netzer:2004,Harrison:2014,Brusa:2015,Carniani:2015,Shen:2016,Bischetti:2017}. The composite spectra in Fig.~\ref{fig:stack_z06} and Fig.~\ref{fig:stack_nir} again shed light on the prevalence of powerful outflows in LoBAL QSOs or otherwise different [\ion{O}{3}] outflow properties. In addition we list the FWHM of the [\ion{O}{3}] line derived from a spectral fit to the composite spectra in Table~\ref{tab:o3width}. We find that the [\ion{O}{3}] profile in the LoBAL QSO composites are largely consistent with the non-BAL QSO composites of comparable luminosity. Thus, we do not see clear evidence of more powerful outflows traced via [\ion{O}{3}] in the LoBAL population. These results are fully consistent with an orientation interpretation of  the LoBAL QSO phenomenon.

\section{Conclusions}  \label{sec:conclusion}
The physical nature of LoBAL QSOs remains poorly understood, with two possible interpretations proposed, an orientation scenario and an evolution scenario. We here present near-IR spectroscopy to study the rest-frame optical properties of 22 luminous LoBAL QSOs at $1.3<z<2.5$ selected from the SDSS to test these different scenarios. We augment this sample with a low-$z$ sample from the literature.
Based on our spectroscopic observations covering the H$\alpha$ and H$\beta$ regions we estimate SMBH masses and Eddington ratios for our sample and generate composite spectra. In addition, we investigate the UV-to-mid-IR SED for our sample. We compare each of these with well matched comparison samples of non-BAL QSOs. Our main results are the following:
\begin{enumerate}
\item We do not find a statistically significant difference in the SMBH masses and Eddington ratios of LoBAL QSOs compared to matched non-BAL QSOs.
\item There are no differences in the UV-to-mid-IR SED apart from dust reddening by $E(B-V)\sim0.14$~dex, most prominent in the UV regime. The similarity in the mid-IR luminosities does not support a large covering fraction of the BALR as implied by the evolution scenario.
\item Our results on the rest-frame optical properties of  LoBAL QSOs remain inconclusive. Overall they are remarkably similar to the general quasar population. While the LoBAL sample at $z<1$ shows strong \ion{Fe}{2} and weak [\ion{O}{3}], we see no enhancement in \ion{Fe}{2} in the two samples at $z>1$ and only the sample at $z\sim2.3$ also shows weak [\ion{O}{3}], while the sample at $z\sim1.5$ has an [\ion{O}{3}] strength broadly consistent with the comparison sample.
\item We do see broad, asymmetric [\ion{O}{3}] line profiles in several cases, indicative of strong ionized outflows. However, we do not find an enhanced prominence of ionized outflow strength in the narrow [\ion{O}{3}] line compared to the general quasar population at similar luminosity.
\end{enumerate}
Overall our results do not provide support for an evolutionary scenario in which LoBAL QSOs represent a young, short-lived AGN phase. They are rather consistent with an orientation interpretation of the LoBAL phenomenon.
LoBALs are not predominantly in an Eddington limited growth phase with on average high Eddington ratio. They are largely indistinguishable from the general quasar population in their rest-frame optical to mid-IR properties. Their \ion{Fe}{2} and [\ion{O}{3}] emission line strengths may point to an intrinsic difference, but to firmly establish if the low-$z$ trends also hold at higher-$z$ requires a larger sample and/or better quality near-IR spectroscopy. 

The crucial tests for the youth scenario of LoBAL QSOs comes likely from their star formation rates and major merger fractions. While the only four LoBAL QSOs at $z<0.4$ support this scenario \citep{Canalizo:2002} it remains to be established if this holds for the broader LoBAL population, in particular at $z>1$, towards the peak of AGN and star formation activity.

\acknowledgments
A.S. is supported by the EACOA fellowship and acknowledges support by JSPS KAKENHI Grant Number 26800098. M.S. acknowledges support by JSPS KAKENHI Grant No. 16H01111. X.-B.Wu thanks the supports by the NSFC grants No.11373008 and 11533001, the National Key Basic Research Program of China 2014CB845700, and from the Ministry of Science and Technology of China under grant 2016YFA0400703.

We thank Toru Misawa for helpful comments and Peng Jiang and Wenjuan Liu for assistance during the Triplespec observations. We thank Ted Boroson for kindly providing us with his iron template for I~Zwicky~1 and Liam Coatman for providing Table~2 from \citet{Coatman:2017}.

Based in part on data collected at Subaru Telescope, which is operated by the National Astronomical Observatory of Japan.
This research uses data obtained through the Telescope Access Program (TAP), which has been funded by the National Astronomical Observatories of China, the Chinese Academy of Sciences (the Strategic Priority Research Program "The Emergence of Cosmological Structures" Grant No. XDB09000000), and the Special Fund for Astronomy from the Ministry of Finance. 
Observations obtained with the Hale Telescope at Palomar Observatory were obtained as part of an agreement between the National Astronomical Observatories, Chinese Academy of Sciences, and the California Institute of Technology.

\begin{figure*}
\centering
\includegraphics[width=17cm,clip]{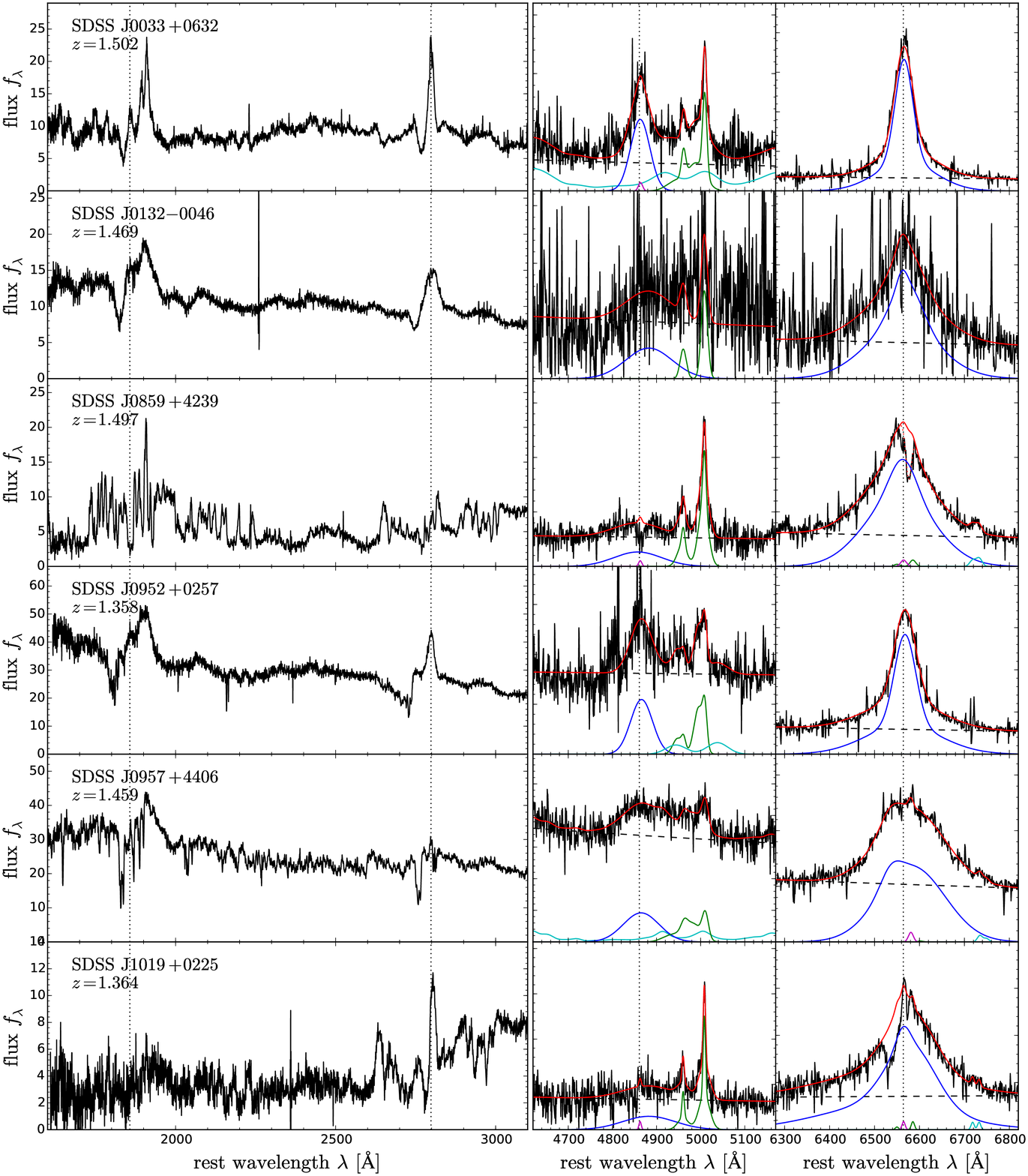} 
\caption{Individual spectra for the first half of the LoBAL sample at $z\sim1.5$. The left panels shows the optical spectra from SDSS-I/II. The location of the \ion{Al}{3} and \ion{Mg}{2} line based on the near-IR redshift is indicated by the vertical dotted lines. The strong absorption lines at \ion{Al}{3} and \ion{Mg}{2} are clearly visible.
The right panels show the H$\beta$ and H$\alpha$ line regions from the near-IR spectra together with our best fit spectral model (red line). The position of H$\beta$ and H$\alpha$ is marked by the vertical dotted line. The model includes a power-law continuum (black dashed line), an \ion{Fe}{2} template (cyan), a multi-Gauss model for the broad Balmer lines (blue) and the [\ion{O}{3}] doublet (green) and a narrow Balmer line (magenta), [\ion{N}{2}] doublet (green) and [\ion{S}{2}] doublet (cyan) component if justified.
}
\label{fig:allspec_z15} 
\end{figure*}

\begin{figure*}
\centering
\includegraphics[width=17cm,clip]{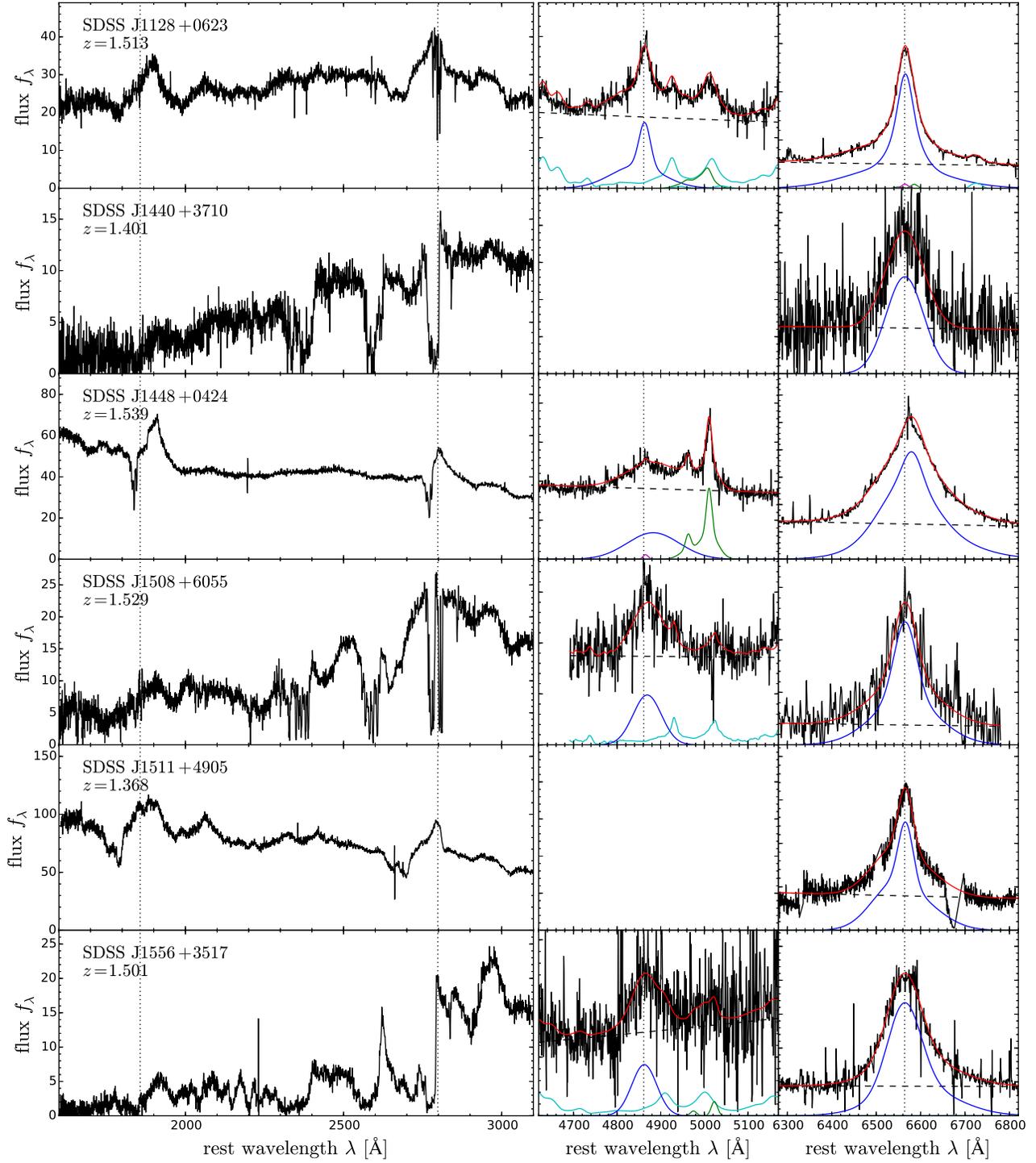} 
\caption{Same as Fig.~\ref{fig:allspec_z15} for the second half of the LoBAL sample at $z\sim1.5$.}
\label{fig:allspec_z15b} 
\end{figure*}

\begin{figure*}
\centering
\includegraphics[width=17cm,clip]{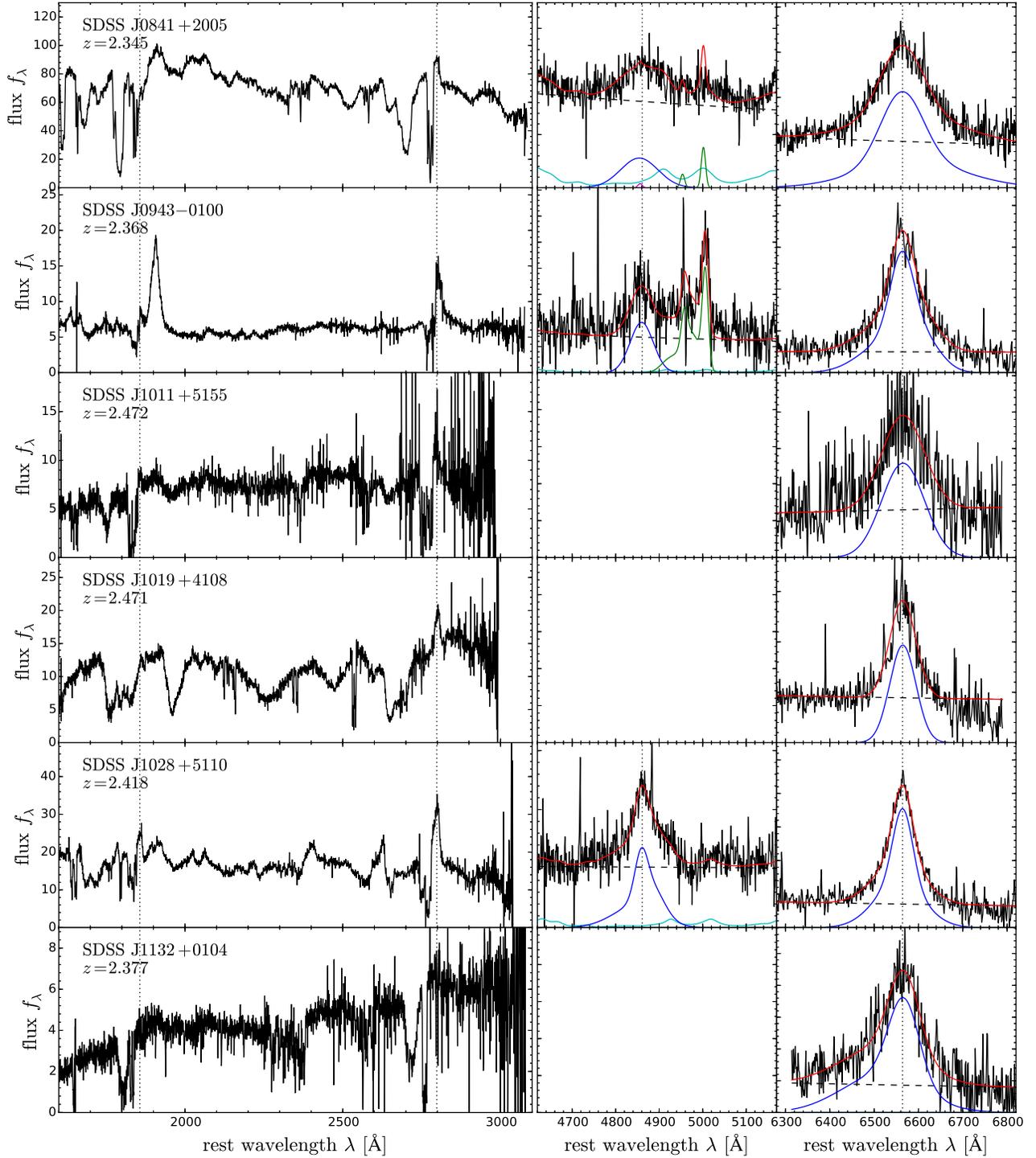} 
\caption{Individual spectra for the first half of the LoBAL sample at $z\sim2.3$. The left panels shows the optical spectra from BOSS with the improved spectrophotometry by \citet{Margala:2016}. The right panels show the H$\beta$ and H$\alpha$ line regions as in Fig.~\ref{fig:allspec_z15}.}
\label{fig:allspec_z23} 
\end{figure*}

\begin{figure*}
\centering
\includegraphics[width=17cm,clip]{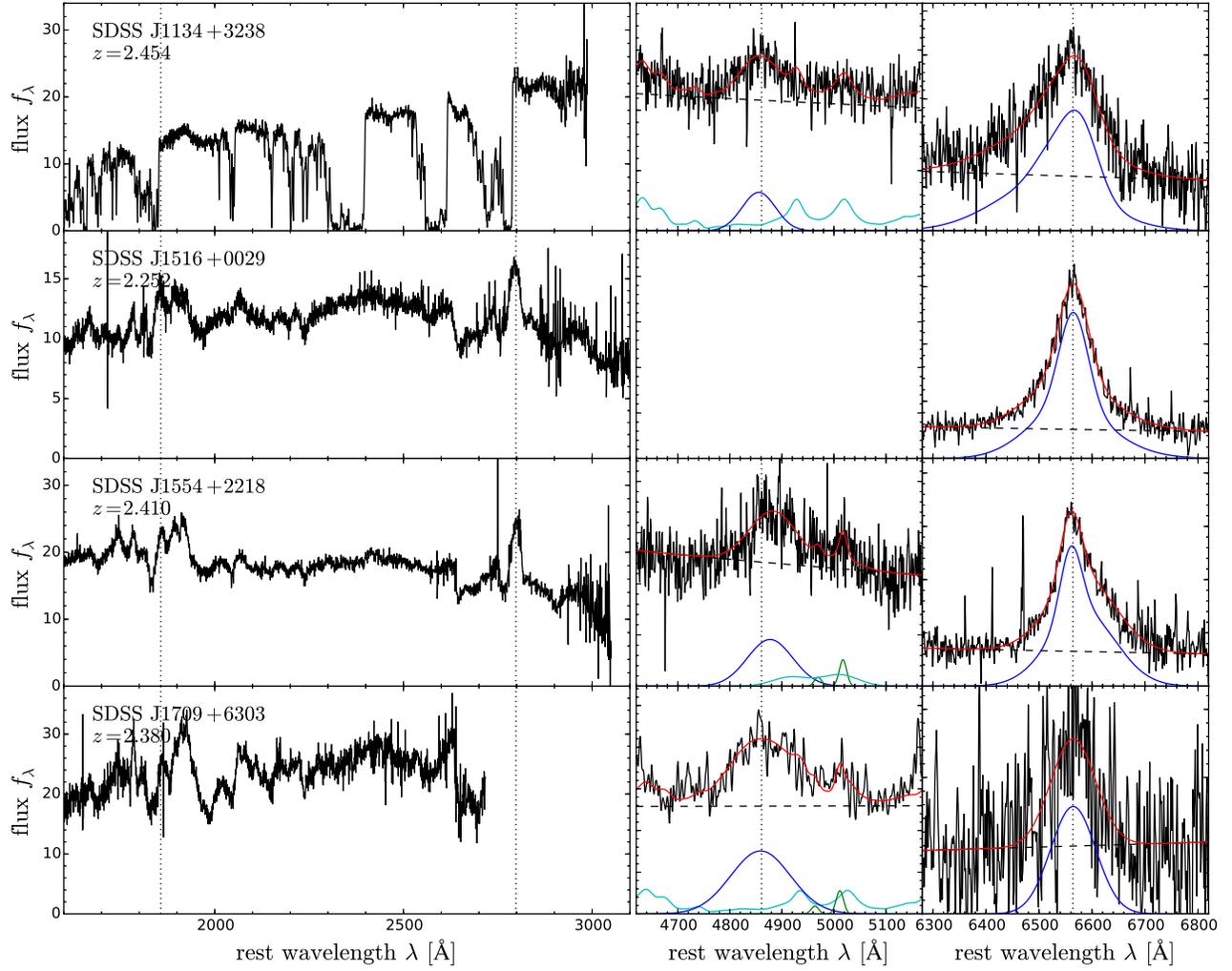} 
\caption{Same as Fig.~\ref{fig:allspec_z23} for the second half of the LoBAL sample at $z\sim2.3$. The optical spectra for SDSS J1709+6303 is based on SDSS-I/II, since it did not have a spectrum from BOSS available.}
\label{fig:allspec_z23b} 
\end{figure*}

\appendix 
We here show the spectra and best spectral model fit for the Balmer line regions for our two LoBAL QSO samples at $z\sim1.5$ (Figs.~\ref{fig:allspec_z15} and \ref{fig:allspec_z15b}) and $z\sim2.3$ (Figs.~\ref{fig:allspec_z23} and \ref{fig:allspec_z23b}).


\clearpage

\end{document}